\documentclass{jpp}

\usepackage{graphicx}
\usepackage{dcolumn}
\usepackage{bm}

\usepackage[utf8]{inputenc}
\usepackage[T1]{fontenc}
\usepackage{etoolbox}
\usepackage{amsmath}

\usepackage{natbib}
\usepackage{caption}
\usepackage{subcaption}
\usepackage{array}
\usepackage{multirow}
\usepackage{bigdelim}
\usepackage{makecell}
\usepackage{nicematrix}
\usepackage{booktabs}
\usepackage{tabularray}
\usepackage{float}
\usepackage{siunitx}
\providecommand{\RadMomflux}[1][s]{\ensuremath{\pi^{(R\tor)}_{#1}}}
\usepackage[capitalise, nameinlink]{cleveref}

\newcommand{\mctrans}{{\texttt{MCTrans++}}}

\providecommand{\grad}{\bm{\nabla}}

\newcommand{\dg}{\cdot\grad}

\newcommand{\vth}[1][s]{\ensuremath{v_{\mathrm{th}_#1}}}

\newcommand{\pd}[2]{\ensuremath{ \frac{\partial #1} {\partial #2} } }

\providecommand{\eqref}[1]{Eq.\ (\ref{#1})}

\providecommand{\Or}[1]{\mathcal{O}#1}

\newcommand{\tor}{\phi} 
\newcommand{\pot}{\varphi} 

\newcommand{\HeatFlux}[1][s]{q_{#1}}
\newcommand{\MomentumFlux}[1][s]{\pi^{(\psi\tor)}_{#1}}

\newcommand{\inertia}{J}

\newcommand{\infrac}[2]{ \left.{#1}\middle/{#2}\right. }

\newcommand{\angvel}{\omega}
\newcommand{\cycfreq}[1][s]{\Omega_{#1}}

\newcommand{\cs}{\ensuremath{c_{\mathrm{s}}}} 

\RequirePackage{amsmath}
\makeatletter
\newcommand*\rel@kern[1]{\kern#1\dimexpr\macc@kerna}
\newcommand*\widebar[1]{%
  \begingroup
  \def\mathaccent##1##2{%
    \rel@kern{0.8}%
    \overline{\rel@kern{-0.8}\macc@nucleus\rel@kern{0.2}}%
    \rel@kern{-0.2}%
  }%
  \macc@depth\@ne
  \let\math@bgroup\@empty \let\math@egroup\macc@set@skewchar
  \mathsurround\z@ \frozen@everymath{\mathgroup\macc@group\relax}%
  \macc@set@skewchar\relax
  \let\mathaccentV\macc@nested@a
  \macc@nested@a\relax111{#1}%
  \endgroup
}
\makeatother

\usepackage[nonumberlist,nopostdot,shortcuts,style=alttree,section=section]{glossaries}

\glssetwidest{\qquad \quad}
\newglossary[slg]{symbols}{sym}{sbl}{Symbols}
\newglossary[sug]{subscripts}{sub}{suo}{Sub- and Superscripts}
\makeglossaries
\newglossaryentry{sub:max}
{
	type = {subscripts},
	name = {$\mathrm{max}$},
	description = {Maximum value}
}

\newglossaryentry{sub:min}
{
	type = {subscripts},
	name = {$\mathrm{min}$},
	description = {Minumum value}
}

\newglossaryentry{sub:exh}
{
	type = {subscripts},
	name = {$\mathrm{exh}$},
	description = {Exhaust}
}

\newglossaryentry{sub:e}
{
	type = {subscripts},
	name = {$e$},
	description = {Electron}
}

\newglossaryentry{sub:i}
{
	type = {subscripts},
	name = {$i$},
	description = {Ion}
}

\newglossaryentry{sub:s}
{
	type = {subscripts},
	name = {$s$},
	description = {Species}
}

\newglossaryentry{sub:n}
{
	type = {subscripts},
	name = {$n$},
	description = {Neutral}
}

\newglossaryentry{sub:alpha}
{
	type = {subscripts},
	name = {$\alpha$},
	description = {Alpha}
}

\newglossaryentry{sub:star}
{
	type = {subscripts},
	name = {$*$},
	description = {Dimensionless}
}

\newglossaryentry{sub:f}
{
	type = {subscripts},
	name = {$f$},
	description = {Fluid}
}

\newglossaryentry{sub:r}
{
	type = {subscripts},
	name = {$r$},
	description = {Relative}
}

\newglossaryentry{sub:th}
{
	type = {subscripts},
	name = {$th$},
	description = {Thermal}
}

\newglossaryentry{sub:eff}
{
	type = {subscripts},
	name = {$\mathrm{eff}$},
	description = {Effective}
}

\newglossaryentry{sub:brem}
{
	type = {subscripts},
	name = {$\mathrm{Brem}$},
	description = {Bremsstrahlung}
}

\newglossaryentry{sub:cyc}
{
	type = {subscripts},
	name = {$\mathrm{cyc}$},
	description = {Cyclotron}
}

\newglossaryentry{sub:thermal}
{
	type = {subscripts},
	name = {$\mathrm{thermal}$},
	description = {Thermal}
}

\newglossaryentry{sub:A}
{
	type = {subscripts},
	name = {$A$},
	description = {Alfv\'en}
}

\newglossaryentry{sym:E}
{
	type = {symbols},
	name = {$E$},
	description = {Electric field [\si{\volt\per\meter}]}
}

\newglossaryentry{sym:B}
{
	type = {symbols},
	name = {$B$},
	description = {Magnetic field [\si{\tesla}]}
}

\newglossaryentry{sym:A}
{
	type = {symbols},
	name = {$A$},
	description = {Vector potential [\si{\tesla \meter}]}
}

\newglossaryentry{sym:a}
{
	type = {symbols},
	name = {$a$},
	description = {Radial length scale, half-width of plasma [\si{\meter}]}
}

\newglossaryentry{sym:L}
{
	type = {symbols},
	name = {$L$},
	description = {Axial length scale [\si{\meter}]}
}

\newglossaryentry{sym:u}
{
	type = {symbols},
	name = {$u$},
	description = {Bulk-flow velocity [\si{\meter\per\second}]}
}

\newglossaryentry{sym:omega}
{
	type = {symbols},
	name = {$\omega$},
	description = {Toroidal angular velocity [\si{\radian\per\second}]}
}

\newglossaryentry{sym:R}
{
	type = {symbols},
	name = {$R$},
	description = {Radial coordinate [\si{\meter}]}
}

\newglossaryentry{sym:tor}
{
	type = {symbols},
	name = {$\tor$},
	description = {Azimuthal coordinate [\si{\radian}]}
}

\newglossaryentry{sym:z}
{
	type = {symbols},
	name = {$z$},
	description = {Axial coordinate [\si{\meter}]}
}

\newglossaryentry{sym:mach}
{
	type = {symbols},
	name = {$M$},
	description = {Mach number}
}

\newglossaryentry{sym:cs}
{
	type = {symbols},
	name = {$\cs$},
	description = {Sound speed (cold ion limit) [\si{\meter \per \second}]}
}

\newglossaryentry{sym:psi}
{
	type = {symbols},
	name = {$\psi$},
	description = {Poloidal flux [\si{\tesla \meter \squared}]}
}

\newglossaryentry{sym:Phi}
{
	type = {symbols},
	name = {$\Phi$},
	description = {Applied potential, see \cref{eq:electricField} [\si{\volt}]}
}

\newglossaryentry{sym:k}
{
	type = {symbols},
	name = {$k$},
	description = {Transparency factor for cyclotron power}
}

\newglossaryentry{sym:varphi}
{
	type = {symbols},
	name = {$\varphi$},
	description = {Ambipolar potential, see \cref{eq:electricField} [\si{\volt}]}
}

\newglossaryentry{sym:e}
{
	type = {symbols},
	name = {$e$},
	description = {Electron charge [\si{\coulomb}]}
}

\newglossaryentry{sym:m}
{
	type = {symbols},
	name = {$m$},
	description = {Mass [\si{\kilogram}]}
}

\newglossaryentry{sym:rho}
{
	type = {symbols},
	name = {$\rho$},
	description = {Gyroradius [\si{\meter}]}
}

\newglossaryentry{sym:n}
{
	type = {symbols},
	name = {$n$},
	description = {Density [\si{\per \meter \cubed}]}
}

\newglossaryentry{sym:T}
{
	type = {symbols},
	name = {$T$},
	description = {Temperature [\si{\electronvolt}]}
}

\newglossaryentry{sym:Gamma}
{
	type = {symbols},
	name = {$\Gamma$},
	description = {Radial particle flux [\si{\per \meter \squared \per \second}]}
}

\newglossaryentry{sym:q}
{
	type = {symbols},
	name = {$q$},
	description = {Radial heat flux [\si{\watt \per \meter \squared}]}
}

\newglossaryentry{sym:J}
{
	type = {symbols},
	name = {$J$},
	description = {Azimuthal moment of interia [\si{\kilogram \per \meter}]}
}

\newglossaryentry{sym:pi}
{
	type = {symbols},
	name = {$\RadMomflux$},
	description = {Radial flux of azimuthal angular momentum [\si{\newton \per \meter}]}
}

\newglossaryentry{sym:jR}
{
	type = {symbols},
	name = {$j_R$},
	description = {Radial current density [A \si{\per \meter \squared}]}
}

\newglossaryentry{sym:Q}
{
	type = {symbols},
	name = {$\dot{Q}$},
	description = {Heating [\si{\watt \per \meter \cubed}]}
}

\newglossaryentry{sym:S}
{
	type = {symbols},
	name = {$S$},
	description = {Source term}
}

\newglossaryentry{sym:IR}
{
	type = {symbols},
	name = {$I_R$},
	description = {Radial current [A]}
}

\newglossaryentry{sym:Z}
{
	type = {symbols},
	name = {$Z$},
	description = {Charge number}
}

\newglossaryentry{sym:Xi}
{
	type = {symbols},
	name = {$\Xi$},
	description = {Potential energy [\si{\joule}]}
}

\newglossaryentry{sym:N}
{
	type = {symbols},
	name = {$N$},
	description = {Arbitrary density, constant along field line [\si{\per \meter \cubed}]}
}

\newglossaryentry{sym:tau}
{
	type = {symbols},
	name = {$\tau$},
	description = {Either confinement or collision time [\si{\second}] or temperature ratio}
}

\newglossaryentry{sym:Rmirror}
{
	type = {symbols},
	name = {$R_\mathrm{mirror}$},
	description = {Mirror ratio}
}

\newglossaryentry{sym:nu}
{
	type = {symbols},
	name = {$\nu$},
	description = {Collision frequency [\si{\per \second}]}
}

\newglossaryentry{sym:v}
{
	type = {symbols},
	name = {$v$},
	description = {Velocity [\si{\meter \per \second}]}
}

\newglossaryentry{sym:V}
{
	type = {symbols},
	name = {$V$},
	description = {Plasma volume [\si{\meter \cubed}]}
}

\newglossaryentry{sym:Omega}
{
	type = {symbols},
	name = {$\Omega$},
	description = {Gyrofrequency [\si{\per \second}]}
}

\newglossaryentry{sym:Rns}
{
	type = {symbols},
	name = {$R_{ns}$},
	description = {Collision rate of neutrals with species $s$ [\si{\per \meter \cubed \per \second}]}
}

\newglossaryentry{sym:sigma}
{
	type = {symbols},
	name = {$\sigma$},
	description = {Reaction cross-section [\si{\meter \squared}]}
}

\newglossaryentry{sym:f}
{
	type = {symbols},
	name = {$f$},
	description = {Distribution function}
}

\newglossaryentry{sym:P}
{
	type = {symbols},
	name = {$P$},
	description = {Power [\si{\watt \per \meter \cubed}]}
}

\newglossaryentry{sym:mu}
{
	type = {symbols},
	name = {$\mu$},
	description = {Magnetic moment [\si{\joule \per \tesla}]}
}

\newglossaryentry{sym:flost}
{
	type = {symbols},
	name = {$f_\mathrm{lost}$},
	description = {Fraction lost}
}

\newglossaryentry{sym:Qsci}
{
	type = {symbols},
	name = {$Q_\mathrm{sci}$},
	description = {Scientific gain}
}

\crefname{equation}{}{}

\title{\mctrans{}: A 0-D Model for Centrifugal Mirrors}

\author{Nick R. Schwartz\aff{1, 2}
	\corresp{\email{nickschw@umd.edu}},
	Ian G. Abel\aff{2},
	Adil B. Hassam\aff{2, 3},
	Myles Kelly\aff{2, 4}
	\and Carlos A. Romero-Talam\'as\aff{2, 5}}

\affiliation{\aff{1}Department of Materials Science and Engineering, University of Maryland, College Park, MD 20742, USA
	\aff{2}Institute for Research in Electronics and Applied Physics, University of Maryland, College Park, MD 20742, USA
	\aff{3}Department of Physics, University of Maryland, College Park, MD 20742, USA
	\aff{4}Department of Mechanical Engineering, University of Maryland, College Park, MD 20742, USA
	\aff{5}Department of Mechanical Engineering, University of Maryland, Baltimore County, MD 21250, USA}

\begin{document}
	\maketitle
	\begin{abstract}
		The centrifugal mirror confinement scheme incorporates supersonic rotation of a plasma into a magnetic mirror device. This concept has been shown experimentally to drastically decrease parallel losses and increase plasma stability as compared to prior axisymmetric mirrors. \mctrans{} is a 0D scoping tool which rapidly models experimental operating points in the Centrifugal Mirror Fusion Experiment (CMFX) at the University of Maryland. In the low-collisionality regime, parallel losses can be modeled analytically. A confining potential is set up that is partially ambipolar and partially centrifugal. Due to the stabilizing effects of flow-shear, the perpendicular losses can be modeled as classical. Radiation losses such as Bremsstrahlung and cyclotron emission are taken into account. A neutrals model is included, and, in some circumstances, charge-exchange losses are found to exceed all other loss mechanisms. 
		We use the SUNDIALS ARKODE library to solve the underlying equations of this model; the resulting software is suitable for scanning large parameter spaces, and can also be used to model time-dependent phenomena such as a capacitive discharge. \mctrans{} has been used to verify results from prior centrifugal mirrors, create an experimental plan for CMFX, and find configurations for future reactor-scale fusion devices.
	\end{abstract}
	\glsaddall
	\printglossaries
	\section{\label{sec:Introduction}Introduction}
	Axisymmetric mirror machines were at one time attractive as thermonuclear fusion devices because of their engineering simplicity, high beta, and steady-state operating capability. However, these devices are plagued by poor axial confinement and the interchange instability \citep{Post1987}. The centrifugal mirror confinement scheme, in contrast, incorporates supersonic rotation of a plasma into a conventional axisymmetric magnetic mirror device. Centrifugal confinement greatly reduces loss rates and is beneficial for removing impurities \citep{Lehnert1971}. Other concepts for improved confinement and stabilization in a magnetic mirror (or ``open trap'') are outlined in \citet{Ryutov2011}. 
	
	Supersonic rotation has been demonstrated to improve axial confinement \citep{Teodorescu2010} and stability \citep{Huang2001}, thereby reducing perpendicular losses too. This concept was first demonstrated with the Ixion device at Los Alamos National Lab \citep{Baker1961, Baker1961_2}, and variations have been constructed at the Institute of Nuclear Physics in Russia (PSP-2 \citep{Volosov2009}) and the University of Maryland (MCX \citep{Ellis2005}), but none of these are currently in operation.
	
	Most recently, construction has been completed and first plasma achieved at the Centrifugal Mirror Fusion Experiment (CMFX) at the University of Maryland \citep{Romero-Talamas2022}. This paper discusses the details and results from \mctrans{}, a 0D scoping tool which is primarily used to model experimental operating conditions in CMFX. Additionally, \mctrans{} can be used to predict the performance of reactor-scale centrifugal mirrors, as well as verify results from centrifugal mirror experiments mentioned previously.
	
	There are significant engineering challenges to developing a fusion power plant based on the centrifugal mirror, but the aim of this paper is to demonstrate what is physically achievable if experimental concerns can be overcome. However, it is worth mentioning that two major challenges are impurity ions sputtered off plasma-facing surfaces and avoiding electrical breakdown from the necessary high voltages. Other research to tackle these issues is ongoing in parallel to this theoretical work.
	
	The rest of this paper is structured as follows: in \cref{sec:Physics Model}, we survey the basis for the underlying physics, including formulae for parallel and perpendicular loss rates. In \cref{sec:Features}, optional features of \mctrans{} are discussed, including time dependence, neutrals and radiation models, and alpha heating. Results and discussion are covered in \cref{sec:Results and Discussion} and conclusions in \cref{sec:Conclusions}. \cref{sec:ShearFlowStabilization} is a literature review of shear flow stabilization, both in centrifugal mirrors and tokamaks.
	
	\section{\label{sec:Physics Model}Physics Model}
\begin{figure}
		\centering
		\includegraphics[width=0.6\textwidth]{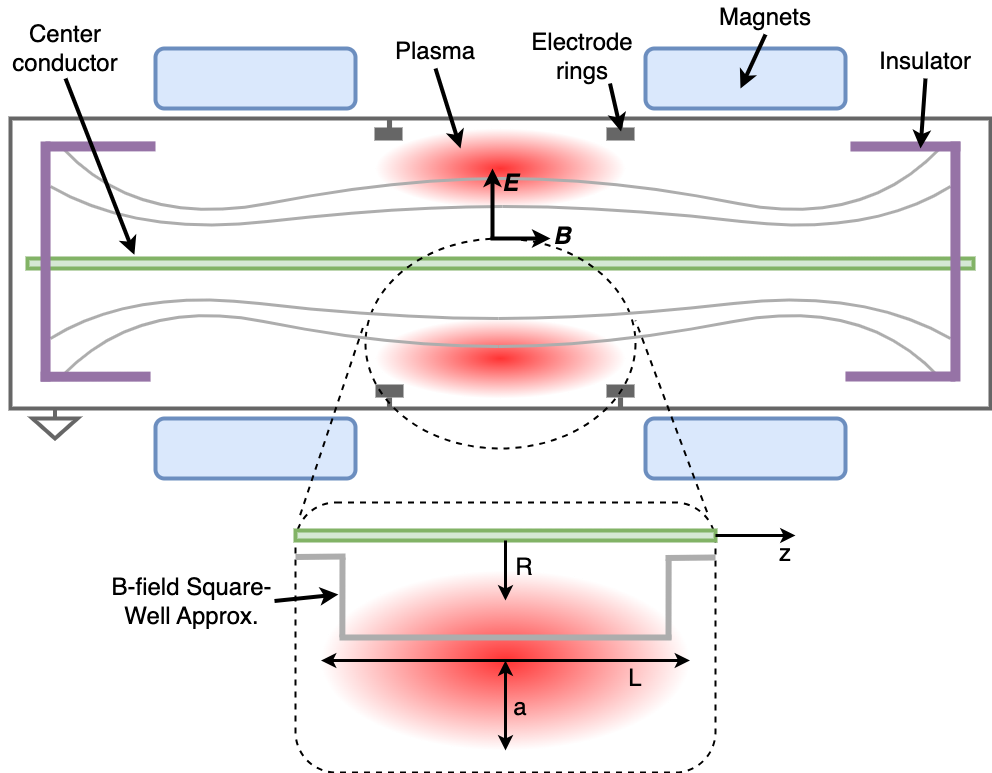}
		\caption{Diagram of CMFX. The axial magnetic field is generated by superconducting magnets, and the electric field by directly biasing the central electrode and grounding the vacuum chamber. Supersonic rotation is achieved by the $\bm{E} \times \bm{B}$ velocity. The magnetic field terminates on electrically insulating surfaces so that the voltage can vary across the field lines.}
		\label{fig:CMFX_diagram}
	\end{figure}

	A diagram of CMFX presents the most important features, namely the magnetic field, central conductor, insulators, and grounded chamber and ring electrodes (\cref{fig:CMFX_diagram}). The central conductor is biased by a large voltage from a capacitor bank, and a large radial electric field generates supersonic $\bm{E} \times \bm{B}$ flows. The potential drop from the center electrode to the grounded ring electrodes is held fixed, and it has been shown previously that the velocity and temperature profiles across the radial width of the plasma are approximately parabolic \citep{Romero-Talamas2012,Huang2001}. We take the transverse length scale $a$ to be half the width, whereas the parallel length scale $L$ is the axial extent of the plasma.
	
	The grounded ring electrodes are an improvement over the designs in other centrifugal mirrors, where the outer radius of the plasma was limited by the chamber wall (as in MCX and Ixion) or a liner coincident with the vacuum field (as in PSP-2). The ring electrodes present a smaller surface area for plasma-material interactions, thereby decreasing sputtered impurities into the core. The inner- and outer-most flux surfaces are those which intersect the central conductor and ring electrode, respectively. We can therefore calculate the width of the plasma $2a$ as the distance between these flux surfaces at the midplane. Additionally, the simplest, and most transparent, geometric approximation of the magnetic field is the ``square well'' model. In this model we assume the field lines to be straight, directed in the $z$-direction, with a step-function shape (see inset in \cref{fig:CMFX_diagram}).
	
	Prior centrifugal mirrors, like PSP-2 \citep{Abdrashitov1991}, and current experiments, like GDT \citep{Beklemishev2010} and TAE's C-2W \citep{Binderbauer2015}, use edge-biasing with concentric ring electrodes to establish a radial electric field profile in the plasma. \citet{Ryutov2011} provides a good theoretical understanding for stabilization via the combination of edge-biasing and ``line-tying''. This is different from the CMFX approach, which has insulating end plates and a central conductor. Because the total potential drop and the boundary conditions are fixed, diffusive transport smooths the voltage into a singly-peaked profile \citep{Romero-Talamas2012,Huang2001}. And while experiments like GDT and C-2W do use some amount of biasing to create shear-flow stabilization, centrifugal mirrors impose much higher voltages to achieve improved confinement.
	
	\mctrans{} solves simplified transport equations for the centrifugally-confined plasma. The underlying model is discussed in the following sections, and it is primarily based on the assumptions that the plasma is well-confined, strongly-magnetized, and has low-collisionality, all of which can be confirmed \textit{a posteriori}.
	
	\subsection{\label{sec:Fundamentals}Fundamentals}
\begin{figure}
		\centering
		\includegraphics[width=0.7\textwidth]{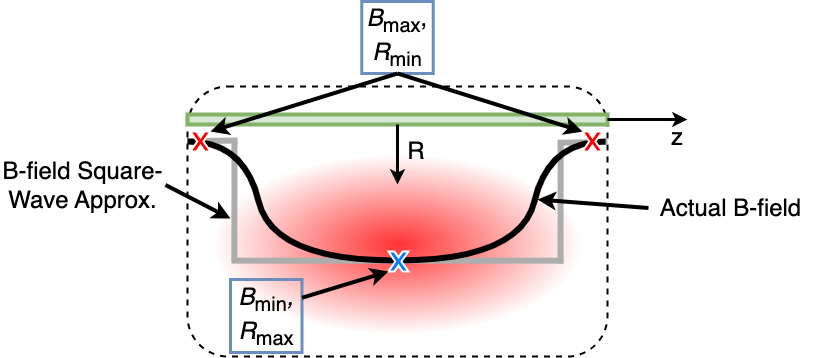}
		\caption{Along a given field line, $R$ falls off along $z$. The blue \textbf{X} indicates the location of $R_\mathrm{max}$ (equivalently $B_\mathrm{min}$) and the red \textbf{X} the location of $R_\mathrm{min}$ (equivalently $B_\mathrm{max}$). This holds true for either the actual magnetic field or the square-well approximation. Thus, according to \cref{confinedDensity}, the density along a flux surface will decrease exponentially as $|z|$ increases because $R$ decreases.}
		\label{fig:potential_drop}
	\end{figure}

	We first present the basic definitions needed to understand a centrifugal mirror. The flow velocity $\bm{u}$ is the same for all species and given by \citep{Ellis2001}
	\begin{equation}
		\bm{u} = \omega R^2 \grad\tor,
		\label{flow_velocity}
	\end{equation}
	where $\omega$ is the azimuthal angular velocity, $R$ is the radius, and $\tor$ the azimuthal coordinate. The Mach number is defined as
	\begin{equation}
		M \equiv \frac{|\bm{u}|}{\cs} = \frac{\omega R_{max}}{c_s} = \omega R_{max} \left/ \sqrt{\frac{T_e}{m_i}} \right. ,
		\label{Mach}
	\end{equation}
	where we take $\cs \equiv \sqrt{\infrac{T_e}{m_i}}$, where $T_e$ and $m_i$ are the electron temperature and ion mass, respectively. We use the subscript `$\mathrm{max}$' to denote the location of maximum radius in the square-well approximation (see \cref{fig:potential_drop}). Note that, typically $T_e \sim T_i$ \citep{Reid2014}, but we take $\cs$ as the cold ion limit of the sound speed, which turns out to be a convenient normalization factor that appears in the derivation of the confining potential (see \cref{sec:Centrifugal Potential}). Additionally, we assume that the magnetic field lies purely in the $R-z$ plane. In cylindrical coordinates $(R,\ \tor,\ z)$, the axisymmetric magnetic field can be expressed in terms of the poloidal flux $\psi$:
	\begin{equation}
		\bm{B} = \grad\psi\times\grad \tor,
	\end{equation}
	where $\psi \equiv R A_\tor$, and $A_\tor$ is the azimuthal component of the vector potential. The electric field is dominantly perpendicular to the magnetic field, so that $\bm{E} \cdot \bm{B} = 0$. To satisfy this condition, we introduce the electric field \citep[also given by (12) in][]{flowtome1} as
	\begin{equation}
		\label{eq:electricField}
		\bm{E} = -\grad \psi \frac{d \Phi}{d\psi} - \grad \pot,
	\end{equation}
	where $\Phi$ is the part of the electrostatic potential associated with the plasma rotation (primarily the applied voltage) and $\pot$ comprises all other pieces of the electrostatic potential.
	
	It is well known that (in the ideal, zero resistivity, limit) the field lines (or equivalently flux surfaces) in a plasma rotate as rigid bodies\footnote{This behavior of isorotation is true when $E_\perp \gg E_\parallel$ \citep{Lehnert1971}. In our configuration, where $T_e \sim T_i$ and $\Phi \gg \varphi$ (see \cref{eq:potential_comparison}), we expect $E_\perp \gg E_\parallel$.} \citep{Ferraro1937, Lehnert1971}. Mathematically, the angular frequency of any such surface $\omega$ obeys
	\begin{equation}
		\bm{B}\dg\omega = 0,
	\end{equation}
	and so $\omega = \omega(\psi)$ is a flux function. In strongly-magnetized plasmas, any part of this flow perpendicular to $\bm{B}$ must be an $\bm{E}\times\bm{B}$ flow and so the angular frequency $\omega$ can be written in terms of the radial derivative of the electric potential $\Phi$ \citep{Lehnert1971}:
	\begin{equation}
		\omega = - \frac{d \Phi}{d\psi},
	\end{equation}
	the radial electric field completely determining the flow profile and vice-versa.
	
	To estimate the size of $\Phi$, we approximate the flow velocity as
	\begin{equation}
		|\bm{u}| \sim \frac{|E|}{|B|} \approx \frac{\Phi}{a |B|} = \frac{e \rho_i \Phi}{a m_i v_{th_i}},
	\end{equation}
	where $e$ is the electron charge, $\rho_i = \infrac{m_i v_{th_i}}{e |B|}$ is the ion gyroradius, and $v_{th_i} = \sqrt{\infrac{2 T_i}{m_i}}$ is the ion thermal velocity. By assuming that the plasma is strongly-magnetized (i.e. the ion gyroradius is small -- $\infrac{\rho_i}{a} \ll 1$) and our rotation is supersonic ($M \gg 1$), and substituting $M \equiv \infrac{|\bm{u}|}{\cs}$, we can compare the size of $\Phi$ and $\pot$, finding
	\begin{equation}
		\Phi \sim M\frac{a}{\rho_i}\frac{T_e}{e} \gg \pot \sim M^2 \frac{T_e}{e}, \label{eq:potential_comparison}
	\end{equation}
	where the Mach number dependence of $\pot$ comes from centrifugal effects (see \cref{ambipolar_pot_simplified}).
	In a centrifugal mirror, the dominant source of the potential $\Phi$ is the voltage applied to the central conductor and $\pot$ is determined by the fact that the plasma must remain quasineutral. 
	For this reason we will often refer to $\pot$ as the ambipolar potential.
	
	If we assume that the plasma rotates at a large Mach number $M \gg 1$, and it has an ion temperature such that the plasma is fully ionized, then the plasma is well-confined and the confinement time is long compared to the collision time (this can be checked \textit{a posteriori}). In such a situation, the plasma is locally Maxwellian (equivalently it is in local thermal equilibrium on a flux surface) and we can write the density $n$ of species $s$ as \citep[see (96) in][]{flowtome1,Catto1998}:
	\begin{equation}
		n_s = N_s (\psi) \exp\left( - \frac{\Xi_s}{T_s} \right) = N_s (\psi) \exp\left( - \frac{Z_s e \pot}{T_s} + \frac{m_s \omega^2 R^2}{2 T_s} \right),
		\label{confinedDensity}
	\end{equation}
	where $N_s$ is an arbitrary flux function with units of \si{\per \meter \cubed}, $\Xi_s$ is the potential energy of a particle (discussed further in \cref{sec:Centrifugal Potential}), and $Z_s$ is the charge number. We see that $n_s$ falls off exponentially along a field line because of its $R$-dependence (see \cref{fig:potential_drop}). Because $\Xi_s$ is a potential, we can choose its zero to be anywhere. We therefore choose $\Xi_s = 0$ at the midplane ($z=0$, $R=R_{\mathrm{max}}$) so that $n_s = N_s$ at the midplane (see \cref{confinedDensity}). This choice simplifies solving for density because we can directly solve for the flux function $N_s$ at the midplane, and then easily compute $n_s$ along a field line.
	
	\mctrans{} primarily solves the transport equations at the midplane, where $N_s = n_s$. \cref{parttrans,heattrans,momtrans}. These have been derived from the low-collisionality, strongly-magnetized transport equations for an axisymmetric rotating plasma. Using the square-well approximation, there is no variation in $z$. The conservation of particles, energy, and azimuthal angular momentum are thus,
	\begin{align}
		\label{parttrans}
		\pd{N_s}{t} + \frac{1}{R} \pd{ }{R} R{\Gamma}_s &= S_{n,s}\\
		\label{heattrans}
		\pd{}{t} \left( \frac{3}{2} N_s T_s \right) + \frac{1}{R} \pd{ }{R} R\HeatFlux &= \RadMomflux \pd{\omega}{R} + Q_s + S_{E,s}\\
		\label{momtrans}
		\pd{}{t}\left( \inertia\omega \right) + \frac{1}{R} \pd{ }{R} R\sum_s\RadMomflux &= - j_R R B + S_{\omega}.
	\end{align}
	In these equations, ${\Gamma}_s$ and ${q}_s$ are the radial particle and heat fluxes of each species, $\inertia = \sum_s m_s N_s R^2$ is the moment of inertia of a flux surface at radius $R$, and $\RadMomflux$ is the radial flux of azimuthal angular momentum. \mctrans{} sets a constant density, but we leave $\pd{N_s}{t}$ in this work to maintain generality. An external power source drives the radial current density $j_R$, which is how the plasma is spun up. 
	The sources of heating in these equations come from viscous heating and the collisional equilibration between species (denoted by $Q_s$).
	We have also included arbitrary sources of particles $S_{n,s}$, energy $S_{E,s}$, and angular momentum $S_{\omega,s}$; these source terms will be used to account for additional energy sources such as alpha-particle heating. Currently, \mctrans{} models a particle source that is feedback-controlled to maintain the electron density at a fixed value. We use the SUNDIALS ARKODE library to solve these conservation equations via a 4th-order implicit Runge-Kutta scheme.
	
	We often wish to consider the behavior of a centrifugal plasma given a fixed input voltage (and thus fixed $\Phi$). In this case, the momentum transport equation is used to determine the radial current drawn from the power supply, $I_R = 2\pi R L j_R$.\footnote{This simplicity is only achieved in the 0D approximation, in general merely knowing the total potential drop across the plasma is not enough information to determine $\omega$ completely.} We never need to solve \cref{momtrans} for $\omega$ under this fixed input. Instead, if we hold the input power or current fixed then \cref{momtrans} would be solved to find $\omega$.
	
	In steady-state operation, \mctrans{} solves the time-dependent equations with internal time-steps until the solution converges to balance heat generation and heat losses. \cref{sec:TimeDependenceTheory} discusses the time-dependent capabilities.
	In the following subsections we will explain the approximations used to calculate those heat losses explicitly in terms of the system state.
	
	\subsection{\label{sec:Parallel Transport}Parallel Transport}
	We begin with a discussion on parallel transport of particles, heat, and momentum to feed into their respective equations \cref{parttrans,heattrans,momtrans}. The confining potential in the parallel direction primarily arises from the centrifugal potential, with a small (but not negligible) component from the electrostatic potential. In the following section, we derive said potential, $\Xi_s$, and then determine parallel losses in \cref{sec:Parallel Losses}.
	
	\subsubsection{\label{sec:Centrifugal Potential}Centrifugal Potential}
	Contributions to the potential energy of a charged particle on a rotating field are two fold: (1) the electrostatic potential $Z_s e \pot$ and (2) the centrifugal potential, which is given by $m_s \omega^2 R^2 / 2$ for a particle of mass $m_s$ at radius $R$ rotating with velocity $\omega$. The potential energy of a particle of species $s$ is then
	\begin{equation}
		\Xi_s = Z_s e \pot - \frac{m_s}{2} \omega^2 R^2.
		\label{eq:particlePotential}
	\end{equation}
	The first term in this potential is confining (i.e. negative) for electrons, but deconfining for ions, whereas the second centrifugal term is confining for all species. We can reuse results from previous work \citep[notably][]{Pastukhov1974} which consider only electrostatic confinement, simply by making the substitution $Z_s e \pot \rightarrow \Xi_s$. Physically, the first term serves to keep the electrons (which are light and barely affected by the centrifugal force) next to the ions, which are pushed to regions of large $R$ by the centrifugal force. Hence, the potential $\pot$ can be found by insisting that the plasma is quasineutral along field lines and that the loss rate is ambipolar \citep{Post2004}.
	
	The ambipolar potential can be found by breaking it down into zeroth- and first-order components, $\pot = \pot_0 + \pot_1$. The leading order term is found by imposing quasineutrality assuming no losses, while the second term (found in \cref{sec:Parallel Losses}) is determined by insisting that the losses preserve quasineutrality.
	
	If we now insist that the plasma is made up of ions (mass $m_i$ and charge $Z_i e$) and electrons (mass $m_e$ and charge $-e$), then the condition for quasineutrality is found by equating \cref{confinedDensity} for ions and electrons
	\begin{equation}
		Z_i N_i(\psi) \exp\left( - \frac{Z_i e\pot_0}{T_i} + \frac{m_i}{2T_i} \omega^2 R^2 \right) = N_e(\psi) \exp\left( \frac{e \pot_0}{T_e} + \frac{m_e}{2T_e} \omega^2 R^2 \right).
		\label{eq:phi0}
	\end{equation}
	We eliminate $N_s$ by noting that $Z_i N_i = N_e$ on a given flux surface $\psi$. And so, up to a possible constant offset, we have
	\begin{equation}
		\pot_0 = \left( \frac{Z_i e}{T_i} + \frac{e}{T_e}\right)^{-1} \left(\frac{m_i}{2 T_i} - \frac{m_e}{2T_e}\right) \omega^2 R^2 \approx \left( \frac{Z_i e}{T_i} + \frac{e}{T_e}\right)^{-1} \frac{m_i}{2 T_i}\omega^2 R^2,
	\end{equation}
	where we assume the temperatures are such that we can drop the term proportional to the electron mass in all further calculations. We've denoted this potential $\pot_0$ because it is $\Or(M^2 T_e/e)$ and is, in fact, the leading order term in an asymptotic series of $\pot$ in $M^{-1}$.\footnote{In computing the densities in \cref{confinedDensity}, we have integrated over a full Maxwellian distribution, neglecting the fact that some small number of high-energy particles are lost along the field line. This is a consistent approximation as we have determined that the potential barrier is $\Or(M^2 T_e/e)$, which to leading order is effectively infinite.} We end with a convenient expression for the potential drop from the center of a flux surface (at $R = R_{max}$) in terms of suitably normalized variables:
	\begin{equation}
		\frac{e \pot_0}{T_e} = \left( Z_i + \tau \right)^{-1} \left( \frac{R^2}{R_{\mathrm{max}}^2} - 1 \right) \frac{M^2}{2} + \Or(1),
		\label{ambipolar_pot}
	\end{equation}
	where $\tau = T_i/T_e$ is the temperature ratio and we have changed the zero of $\pot_0$ so that it vanishes on midplane. And as a consequence of flux conservation, one can approximate the ratio of radii in terms of the mirror ratio as follows
	\begin{equation}
		\frac{1}{R_{\mathrm{mirror}}} \equiv \frac{B_{\mathrm{min}}}{B_{\mathrm{max}}} \approx \left(\frac{R_{\mathrm{min}}}{R_{\mathrm{max}}}\right)^2, 
	\end{equation}
	and so we can relate the ratio of the radius of the flux surface at throat, $R_\mathrm{min}$, and in the central cell, $R_\mathrm{max}$, to the mirror ratio, $R_{\mathrm{mirror}}$, given by the ratio of magnetic field strengths. This approximation is satisfactory for the vacuum field, but high speed rotation creates a self-consistent field that provides better confinement \citep{Abel2022}.
	
	In the simplest form of \cref{ambipolar_pot}, where $Z_i = 1$, $T_i = T_e$, and $R = R_{\mathrm{min}}$, we end up with 
	\begin{equation}
		\frac{e \pot_0}{T_e} = \left( \frac{1}{R_{\mathrm{mirror}}} - 1 \right) \frac{M^2}{4} + \Or(1),
		\label{ambipolar_pot_simplified}
	\end{equation}
	with the usual $M^2/4$ scaling for the potential drop \citep{Ellis2001}. We leave $Z_i$ as a free parameter elsewhere, but use $Z_i = 1$ simply to show agreement with \citet{Ellis2001}. We will see in \cref{sec:Parallel Losses} that this scaling is very important for minimizing parallel losses because it appears in an exponential term for the loss rate.
	
	Equation \cref{ambipolar_pot} is only the leading-order term in the $M \gg 1$ expansion of $\pot$, and so we collect all higher-order terms and denote them by $\pot_1$. To compute the next-order terms in this series we need to know the particle loss rate and hence parallel transport (discussed in \cref{sec:Parallel Losses}). For electrons, the only term in the potential energy is the electrostatic potential:
	\begin{equation}
		\label{eq:electron_potential}
		\begin{split}
			\Xi_e &= \frac{1}{Z_i + \tau} \left( 1 - \frac{R^2}{R_{\mathrm{max}}^2} \right) \frac{M^2}{2} T_e - e \pot_1,
		\end{split}
	\end{equation}
	but for ions we need to include the centrifugal potential to obtain
	\begin{equation}
		\label{eq:ion_potential}
		\begin{split}
			\Xi_i &= \frac{Z_i}{Z_i + \tau} \left( \frac{R^2}{R_{\mathrm{max}}^2} - 1 \right) \frac{M^2}{2} T_e + Z_i e \pot_1 - \frac{m_i}{2} \omega^2 (R^2 - R_{\mathrm{max}}^2) \\
			&= \left[ \frac{Z_i}{Z_i + \tau} \left( \frac{R^2}{R_{\mathrm{max}}^2} - 1 \right) - \left( \frac{R^2}{R_{\mathrm{max}}^2} - 1 \right) \right] \frac{M^2}{2} T_e + Z_i e \pot_1\\
			&= \frac{\tau}{Z_i + \tau} \left( 1 - \frac{R^2}{R_{\mathrm{max}}^2} \right) \frac{M^2}{2} T_e + Z_i e \pot_1,
		\end{split}
	\end{equation}
	Again, $\pot_1$ is the higher-order part of $\pot$ which comes from enforcing ambipolar losses, discussed in the following section.
	
	\subsection{\label{sec:Parallel Losses}Parallel Losses}
	We assume that the time between particle collisions is long compared to the time a particle takes to travel along the mirror machine. This ``low-collisionality'' assumption is true if the plasma is sufficiently hot with sufficiently low density. For reactor-grade plasmas this ratio, called $\nu^* = \left.\nu_{ii} L \right/ \vth$, can be as small as \num{1e-5}, but the assumption holds even in warm plasmas with temperatures of only 100 \si{\electronvolt}. The collisionality parameter $\nu^*$ is an output of the code and validates the assumption \textit{a posteriori} if it is significantly less than one.
	
	The parallel losses are derived from Pastukov's work on low-collisionality plasmas taken in the case of a tandem mirror with an electron-confining electrostatic potential \citep{Pastukhov1974}. Although originally found for electrons alone, we expand the results for a multispecies plasma. \citet{Catto1981} found the parallel loss rates for a generic field shape (Eqn. (40) of that work), and by taking the square-well limit, we obtain the particle loss rate in our notation\footnote{This equation only agrees with the results of \citep{Pastukhov1974} and \citep{Cohen1978} in the limit of $R_{\mathrm{mirror}} \rightarrow \infty$. We've chosen the formula in \citep{Catto1981} as opposed to that in these two works because it is more pessimistic (i.e. has higher loss rates). The error induced by not knowing this prefactor accurately is much smaller than the approximation by a square well, and is comparable with the effect of several other approximations made in this work.}:
	\begin{equation}
		\left(\pd{n_s}{t}\right)_{\mathrm{End Losses}} = 
		- \left( \frac{2 n_s \Sigma}{\sqrt{\pi}} \right) \nu_s \frac{1}{\ln\left( R_{\mathrm{mirror}} \Sigma\right)} \frac{\exp\left( -\Xi_s/T_s \right)}{\Xi_s/T_s},
		\label{eq:ParticleLossRate}
	\end{equation}
	where $\Sigma = Z_i + 1$ for electrons and $\Sigma = 1$ for ions, and $\nu_s$ is the appropriate collision frequency for species $s$ \citep[see (2.5) in][]{Braginskii1965}.
	
	The parallel heat losses can be calculated by multiplying the loss rate in \cref{eq:ParticleLossRate} by the energy of a single particle. We approximate the total energy as the thermal and potential energy, $T_s + \Xi_s$. If $M \gg 1$, $\Xi_s \gg T_s$ following similar logic to \cref{eq:potential_comparison}. 
	
	Similarly, the parallel loss contribution to the total azimuthal angular momentum losses are due to the angular momentum of each particle $m_i \omega R^2$ when it is lost (with electrons carrying negligible momentum). As will be seen in \cref{sec:Power Losses}, this loss mechanism can be dominant. Because flux surfaces rotate rigidly, a particle will have less angular momentum when it is lost farther away from the midplane (i.e. at lower values of $R$, see \cref{fig:potential_drop}). The choice of the exhaust radius, $R_\mathrm{exh}$, for \mctrans{} is therefore critical -- the pessimistic assumption would be to assume that the ion is lost with angular momentum at the midplane ($R = R_\mathrm{max}$); whereas the optimistic assumption would be at the throat ($R = R_\mathrm{min}$). In reality, the momentum is lost when the density has dropped to the point where electrons can no longer shield parallel electric fields, i.e. $B \dg \omega \neq 0$ \footnote{Precisely determining the per-particle loss of angular momentum requires a detailed kinetic study, which will be the subject of future work.}. For this study, we have chosen $R_\mathrm{exh} = R_\mathrm{min}$, but \cref{sec:Power Losses} discusses how varying $R_\mathrm{exh}$ changes results.
	
	In \cref{eq:ParticleLossRate} we see that $\Xi_s$ appears in the exponential. The leading order part of $\Xi_s$ is $\Or(M^2 T_s)$, so this exponential is what strongly suppresses the collisional loss rate. Expanding this exponential, we have
	\begin{equation}
		\exp\left( -\frac{\Xi_s}{T_s} \right) \sim \exp\left[ - (\dots) M^2 \right] \exp\left( -\frac{Z_s e \pot_1}{T_s} \right),
	\end{equation}
	where $(\dots)$ represent the prefactor in \cref{eq:electron_potential,eq:ion_potential} (which are independent of Mach number), and we see that even though $\pot_1$ is small compared to the leading-order part of the potential it has an $\Or(1)$ effect on the loss rate and must be taken into account. \mctrans{} does not separate $\pot_0$ and $\pot_1$, and instead it solves the nonlinear equation for quasineutrality for $\pot_1$ with the initial guess as $\pot_0$, and $\pot=\pot_0+\pot_1$.
	
	We find $\pot_1$ through a root-finding method by equating the electron and ion loss rates along the field line to enforce zero net charge loss. However, at low temperatures ($\lesssim$50 eV), \cref{eq:ParticleLossRate} becomes very sensitive to changes in $T_s$, and produces poor confinement. \mctrans{} may be unable to find an equilibrium for these cases because quasineutrality cannot be satisfied\footnote{We choose a bracketing root-solving method to solve for $\pot$. At low temperatures, no reasonable set of brackets includes the root. Additionally, \mctrans{} is intended for low-collisionality plasmas, where temperatures are typically $\gtrsim$50 eV.}. However, solutions may exist $\lesssim$50 eV and should be checked \textit{a-posteriori} for low collisionality ($\nu^* = \nu_{ii} L_\parallel / \cs \ll 1$). Note that all plots in the Results section (\cref{sec:Results and Discussion}) were checked and have low-collisionality.
	
	\subsection{\label{sec:Perpendicular Transport}Perpendicular Transport}
	
	We now proceed to determine particle, heat, and momentum losses in the perpendicular direction. We make the assumption that turbulent transport will be fully suppressed by the flow shear \citep{Huang2001}, given that the velocity is everywhere perpendicular to the magnetic flux surfaces. A discussion of the literature on the suppression of turbulence by flow shear is given in \cref{sec:ShearFlowStabilization}. We thus assume that the only contributions to these fluxes are the classical collisional fluxes. 
	These can be evaluated from the formulae in \citet{Braginskii1965} or \citet{Helander2002}.

	We begin by considering particle transport. Assuming that the plasma contains no impurities and has $Z_i = 1$, quasineutrality implies that the particle flux of ions must match that of electrons. Therefore (c.f. equation (5.6) of \citet{Helander2002})
	\begin{equation}
	\Gamma_i = \Gamma_e = -D_e N_e \left[ \left( 1 + \frac{T_i}{Z T_e} \right) \frac{1}{N_e} \pd{N_e}{R} - \frac{1}{2T_e} \pd{T_e}{R} + \frac{1}{Z T_e} \pd{T_i}{R} \right],
	\end{equation}
	where the classical electron diffusion rate $D_e$ is given by
	\begin{equation}
	D_e = \frac{T_e}{m_e \cycfreq[e]^2\tau_e},
	\end{equation}
	and $\cycfreq[e]$ is the electron cyclotron frequency, and $\tau_e$ is the electron-electron collision time.

	We take the ion and electron collision times from \citet{Braginskii1965} and convert them to SI units:
	\begin{equation}
		\label{eq:collision_time}
		\tau_i = \frac{6 \sqrt{2 m_i} \pi^{3/2} T_i^{3/2} \epsilon_0^2}{n_i Z^4 e^4 \lambda_i}; \qquad
		\tau_e = \frac{6 \sqrt{2 m_e} \pi^{3/2} T_e^{3/2} \epsilon_0^2}{n_i Z^2 e^4 \lambda_e},
	\end{equation}
	where $\lambda_i$ and $\lambda_e$ are the coulomb logarithms for ions and electrons, taken from \citet{nrlformulary}. These definitions differ by $\sqrt{2}$ from some other definitions of $\tau_s$. 
	
	To completely evaluate the transport equations we need expressions for $\HeatFlux$ and $\RadMomflux$. Similar to the particle flux, we only need to consider the classical collisional heat fluxes. The appropriate form of these fluxes is given by \citet{Braginskii1965}. For the ion heat flux, we use (2.14) and (2.16) from that work:
	\begin{equation}
		\HeatFlux[i] = - 2 \frac{n_i T_i}{m_i \cycfreq[i]^2 \tau_i} \frac{d T_i}{dR},
	\end{equation}
	under the simplifying assumption that $\cycfreq[i] \tau_i \gg 1$ (as appropriate for our plasmas). 
	As we do not evolve the radial temperature profile, we estimate this term by assuming that all the plasma profiles vary on a scale $a$, the half-width of the plasma (\cref{fig:CMFX_diagram}). Thus, we have that
	\begin{equation}
		\frac{1}{R} \pd{ }{R} R\HeatFlux[i] \approx 2\frac{n_i T_i^2}{m_i \cycfreq[i]^2 \tau_i a^2}.
	\end{equation}
	For electrons, (2.13) from \citet{Braginskii1965} gives
	\begin{equation}
		\frac{1}{R} \pd{ }{R} R\HeatFlux[e] \approx \frac{4.66 n_e T_e^2}{m_e \cycfreq[e]^2 \tau_e a^2},
		\label{braginskiiPerpConductivity}
	\end{equation}
	which is approximately $\sqrt{m_e/m_i}$ smaller than the ion heat loss, and usually small (though it is included in \mctrans{}, nonetheless). 
	The numerical coefficient in \cref{braginskiiPerpConductivity} is $Z_i$-dependent and takes the value given for $Z_i = 1$ \citep[for the $Z_i$-dependence of this coefficient, see][Table 1]{Braginskii1965}. We leave $Z_i$ as a free parameter elsewhere. At the time of writing, $Z_i$-dependence of this coefficient is not implemented in \mctrans{}.
	To these conductive heat fluxes, we add $(3/2)n_s T_s \Gamma_s$ to account for the non-zero particle flux. 
	
	To find the radial flux of angular momentum, $\RadMomflux[s]$, we only compute the stress tensor for the ions because electron perpendicular viscosity is at least $m_e/m_i$ smaller than the ion viscosity and thus negligible. Similarly, convective transport of angular momentum by the ions is small compared to their viscous stress. Under these assumptions, from (2.23) of \citep{Braginskii1965}, we obtain
	
	\begin{equation}
		\RadMomflux[i] = \frac{\MomentumFlux[i]}{BR} = - \frac{3}{10} \frac{n_i T_i}{\cycfreq[i]^2 \tau_i} R^2\frac{d\angvel}{dR} \approx- \frac{3}{10} \frac{n_i T_i}{\cycfreq[i]^2 \tau_i} \frac{R^2\angvel}{a}.
		\label{eq:APIonViscosity1}
	\end{equation}
	
	\subsection{Assumptions and Scope}
	Our model is valid when certain assumptions are met, some of which have been discussed already. The plasma must
	\newpage
	\begin{itemize}
		\item be strongly-magnetized,
		\item have low-collisionality,
		\item have a large applied bias in comparison to the ambipolar potential,
		\item have a large mirror ratio,
		\item rotate supersonically. 
	\end{itemize}
	
	\begin{table}
		\begin{center}
			\SetTblrInner{colsep=2pt}
			\begin{tblr}{colspec = {l|ccc||cc|},
					vlines = {2-13}{solid}}
				\cline{2-6}
				& \SetCell[c=3]{c} \textbf{Prior} & & & \SetCell[c=2]{c} \textbf{Predictive} & \\
				\hline
				\textbf{Parameter} & \textbf{Ixion} & \textbf{PSP-2} & \textbf{MCX} & \textbf{CMFX} & \textbf{Reactor} \\
				\hline
				Strongly-magnetized ($\rho^* = \frac{\rho_i}{a} \ll 1$) & 0.024 & 0.035 & 0.052 & 0.166 & 0.091 \\
				Low-collisionality ($\nu^* = \frac{\nu_{ii} L_\parallel}{\cs} \ll 1$) & 815 & \num{1.7e-5} & 6.3 & \num{1.5e-4} & \num{6.8e-5} \\
				Large Mirror Ratio ($R_{\mathrm{mirror}} \gg 1$) & 2.2 & 2.4 & 7.3 & 8.8 & 6.0 \\
				Supersonic ($M \gg 1$) & 2.0 & 6.2 & 2.6 & 5.3 & 4.5 \\
				Plasma $\beta$ & 0.080 & \num{2.5e-4} & 1.0 & 0.11 & 0.27 \\
				Triple Product (\si{\kilo\electronvolt \second \per \meter \cubed}) & \num{2.7e16} & \num{3.0e13} & \num{1.0e16} & \num{2.1e18} & \num{5.9e21} \\
				Plasma width $2a$ (\si{\centi\meter}) & 7.2 & 19 & 20 & 16 & 25 \\
				Plasma length $L$ (\si{\meter}) & 0.38 & 0.4 & 1.4 & 0.6 & 20.0 \\
				$B_{\mathrm{min}}$ (\si{\tesla}) & 0.95 & 0.99 & 0.23 & 0.34 & 3.0 \\
				$\Phi$ (\si{\kilo\volt}) & 7.5 & 360 & 10 & 100 & 5000 \\
				Operating Gas & D & H & H & H & D-T \\
				\hline
			\end{tblr}
			\caption{Parameters (and their corresponding assumptions made in \mctrans{}, if applicable) for prior experiments and projected conditions for CMFX and a reactor scenario. Also included are relevant plasma and experimental parameters for each experiment. Results for Ixion \citep{Baker1961}, PSP-2 \citep{Volosov2009}, and MCX \citep{Teodorescu2008,Reid2014} are from prior experiments, respectively, whereas those for CMFX and Reactor came from \mctrans{} predictive models (see \cref{tab:device_parameters} for input parameters).}
			\label{tab:operating_conditions}
		\end{center}
	\end{table}
	Low-collisionality plasmas are required in \mctrans{} because in the collisional regime, it can no longer be assumed that the temperature is constant on a flux surface. Additionally, the calculations of parallel transport by \citet{Pastukhov1974} and \citet{Catto1981} are not valid for collisional plasmas. This applies to experiments like Ixion and MCX, or to the startup phase when the temperature is low and $\nu^* \gtrsim 1$. Work is in progress to implement a collisional approach. 
	The values to determine these operating conditions for a variety of devices are given in \cref{tab:operating_conditions}, along with a other relevant plasma and experimental parameters. Details of prior experiments are given in \cref{sec:Prior Experiments}, and the results from \mctrans{} in \cref{sec:Scaling}.
	
	\section{\label{sec:Features}Features}
	Our model contains several features, which can be turned ``on'' or ``off'' in \mctrans{}. Typically \mctrans{} is operated in a steady-state mode, but \cref{parttrans,heattrans,momtrans} can be solved in a time-dependent mode. Additionally, models of neutral particles, continuum radiation sources, and alpha heating are provided.
	
	\subsection{\label{sec:TimeDependenceTheory}Time Dependence}
\begin{figure}
		\centering
		\includegraphics[width=0.6\textwidth]{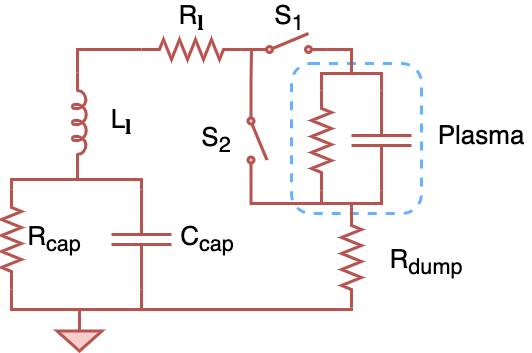}
		\caption{Circuit model of CMFX. $C_{cap}$ and $R_{cap}$ are the capacitance and internal resistance of the capacitor bank, respectively. $R_l$ and $L_l$ are the line resistance and inductance, respectively. The plasma can be modeled as a variable resistor and capacitor in parallel. The dump resistor is in series to the plasma, and when the crowbar is switched, it is assumed all the stored energy from the plasma is transmitted through $R_{dump}$ to ground.}
		\label{fig:circuit_model}
	\end{figure}

	The system can be modeled as a circuit, where the plasma is charged by a capacitor bank and can be discharged through a crowbar into a dump resistor (\cref{fig:circuit_model}). All of the passive circuit elements are static, but the plasma can be thought of as a variable resistor and capacitor in parallel.
	
	The plasma has a voltage-dependent resistance because as it heats up, the current needed to rotate decreases, and thus the effective resistance increases. This is perhaps counter-intuitive as plasma resistance normally decreases with temperature in other devices like tokamaks; but in this case, the resistance increases with velocity shear because the potential gradient between neighboring flux surfaces increases. Additionally, the plasma can be thought of as a capacitor that stores energy in its rotational momentum \citep{Anderson1959}.
	
	Discharges in CMFX proceed as follows:
	\begin{enumerate}
		\item The capacitor bank is charged to some nominal voltage.
		\item Neutral gas is puffed into the chamber.
		\item After some specified time, S1 closes and the capacitor bank discharges, applying high voltage to the central conductor.
		\item A low-temperature plasma forms. The voltage on the capacitors drops because the current draw in this phase is relatively large.
		\item Rotational shear begins to heat up the plasma, and the voltage across the plasma reaches a quasi-steady-state with low current draw.
		\item The circuit is crowbarred by S2, and stored energy in both the capacitor bank and plasma are discharged into a dump resistor.
	\end{enumerate}
	Our model differs from discharges in CMFX in two ways. First, we assume that the electron density is constant, when really the density is time-dependent, especially as the plasma heats up. Second, a starting voltage must be specified, and it should be sufficiently large enough so that the plasma is not highly collisional (effectively skipping Step (iv)). However, \mctrans{} does have the ability to model the crowbar sequence (Step (vi)).
	
	The high voltage circuit can be modeled with simple elements as in \cref{fig:circuit_model}. The equations for voltage and current across the plasma are as follows:
	\begin{align}
		\frac{\mathrm{d}V}{\mathrm{d}t} &= - \frac{I_\mathrm{cap}}{C_\mathrm{cap}} - \frac{V_\mathrm{cap}}{R_\mathrm{cap} C_\mathrm{cap}} \\
		\frac{\mathrm{d}I}{\mathrm{d}t} &= \frac{V_\mathrm{cap} - V - R_l I}{L_l}.
	\end{align}
	As a general rule of thumb, discharges with higher bank capacitance are able to sustain higher voltages across the plasma. As will be seen in \cref{sec:Scaling}, higher plasma voltages almost always produce better performing plasmas.
	
	\subsection{\label{sec:Neutrals Model}Neutrals Model}
	
	\citet{Ng2007} studied neutral penetration into centrifugal mirrors along the axis, finding that the neutral density drops exponentially along the field lines with good centrifugal confinement. However, even a neutral density that is orders of magnitude smaller than the plasma density can have a large effect on power losses (see \cref{sec:Power Losses}), so neutrals cannot be ignored. This section describes the basic model used to determine the neutral density.
	
	To maintain a constant plasma density, neutrals must be supplied to the plasma at the same rate electrons are lost. We assume that a gas puff system provides an ambient source of cold neutrals. To calculate neutral density, we divide the electron loss rate (the sum of parallel and perpendicular losses) by the total ionization rate. Charge exchange is another important loss mechanism for ions. Therefore, we consider three neutral collisional processes: ion- and electron-impact ionization, and charge exchange.
	
	Cross sections for these collisions come from \citet{Janev1993}. Radiative recombination is negligible in the temperature range of interest. The plasma is assumed to be in coronal equilibrium (i.e. the atomic excitation frequency is much smaller than the deexcitation frequency), so excited states are not considered \citep{Drawin1976,Tallents2018}. We also do not consider wall recycling because, although it may decrease the necessary neutral source rate, it does not affect the steady-state neutral density in a 0-D model like \mctrans{}.
	
	We must first calculate the collision rates between neutrals $n$ and some species $s$ per unit volume, given by 
	\begin{equation}
		R_{ns} (\bm{v_s},\ \bm{v_n}) = n_s n_n v_r \sigma(v_r) f_s(\bm{v_s}) f_n(\bm{v_n}) \mathrm{d}^3 \bm{v_s} \mathrm{d}^3 \bm{v_n},
	\end{equation}
	where $n_s$ and $n_n$ are the density of charged species and neutrals, respectively, and $v_r$ is relative velocity between species, $f_s$ is an arbitrary distribution function (normalized such that $\int f_s(\bm{v_s}) \mathrm{d}^3\bm{v_s} = 1$, as is standard in atomic physics), and $\sigma$ is a collision cross section. We choose a Maxwellian distribution for a rotating plasma with a bulk fluid velocity of $|\bm{u}| \equiv M \cs = \omega R_{max}$ such that
	\begin{equation}
		f_s(\bm{v_s}) = \left( \frac{m_s}{2 \pi T_s} \right)^{\frac{3}{2}} \exp \left( -\frac{m_s (\bm{v_s} - \bm{u})^2}{2 T_s} \right),
	\end{equation}
	and the distribution function for the cold neutrals is
	\begin{equation}
		f_n(\bm{v_n}) = \delta( \bm{v_n} ),
	\end{equation}
	where $\delta$ is the Dirac delta function.
	We then transform into spherical coordinates \citep[for the sake of simplifying the integral, as is done in][]{Appelbe2011} where the Jacobian is $\mathrm{d}^3 \bm{v_s} = v_s^2 \sin \theta_s \mathrm{d} v_s \mathrm{d} \theta_s \mathrm{d} \phi_s$. In order to perform the integration, we choose a coordinate system that is local to a single particle and align the z-axis of the transform with the fluid velocity. To be clear, this choice of coordinate system is \textit{not} the global cylindrical coordinate system, where fluid flow is in the azimuthal direction:
	\begin{align*}
		v_{sx} &= v_s \sin \theta_s \cos \phi_s, \quad &u_x &= 0 \\
		v_{sy} &= v_s \sin \theta_s \sin \phi_s, \quad &u_y &= 0 \\
		v_{sz} &= v_s \cos \theta_s, \quad &u_z &= u,
	\end{align*}
	so that the integrand now becomes
	\begin{equation}
		\begin{split}
			R_{ns} (v_s) = &n_s n_n \left( \frac{m_s}{2 \pi T_s} \right)^{\frac{3}{2}} v_s^3 \sigma(v_s) \sin \theta_s \\ 
			&\quad \times \exp \left[ -\frac{m_s (v_s^2 + u^2 - 2 v_s u \cos \theta_s)}{2 T_s} \right] \delta( \bm{v_n} ) \mathrm{d} v_s \mathrm{d} \theta_s \mathrm{d} \phi_s \mathrm{d}^3 \bm{v_n}.
		\end{split}
	\end{equation}
	
	Only the delta distribution is a function of $\bm{v_n}$, so that term integrates out to 1. Completing the integral and taking the thermal velocity to be $\vth = \sqrt{2 T_s / m_s}$, we have
	\begin{equation}
		\begin{split}
			R_{ns} = &\frac{2 n_s n_n}{u \vth \sqrt{\pi}} \exp \left( -\frac{u^2}{\vth^2} \right) \\
			&\quad \times \int_0^\infty v_s^2 \sigma(v_s) \sinh \left( \frac{2 v_s u}{\vth^2} \right) \exp \left( -\frac{v_s^2}{\vth^2} \right) \mathrm{d} v_s.
		\end{split}
	\end{equation}
	To simplify, we expand the $\sinh(\dots)$ term and define a thermal Mach number $M_{th_s} = u / \vth = M \cs / \vth$ such that
	\begin{equation}
		\begin{split}
			R_{ns} = &\frac{n_s n_n}{M_{th_s} \vth^2 \sqrt{\pi}} \int_0^\infty v_s^2 \sigma(v_s) \\
			&\quad \times \left\lbrace \exp \left[ -\left( M_{th_s} - \frac{v_s}{\vth} \right)^2 \right] - \exp \left[ -\left( M_{th_s} + \frac{v_s}{\vth} \right)^2 \right] \right\rbrace \mathrm{d} v_s.
		\end{split}
	\end{equation}
	
\begin{figure}
		\centering
		\includegraphics[width=0.6\textwidth]{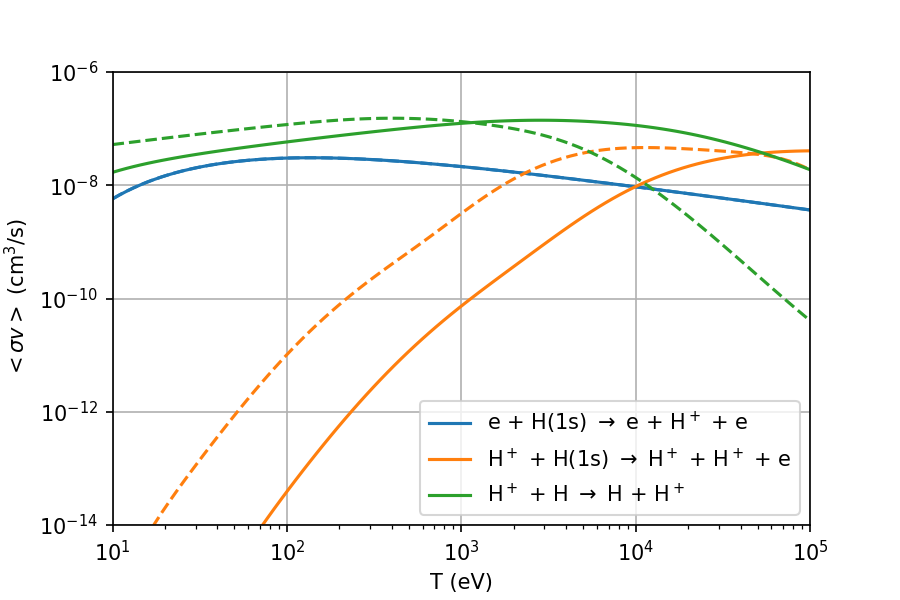}
		\caption{Rate coefficients for a number of collisions involving neutrals. Solid lines are for a non-rotating plasma, while dashed lines are for a plasma with $M=4$. The electron-impact ionization rate is not affected by rotation, and in the limit $M \rightarrow 0$, the dashed lines equal the solid.}
		\label{fig:ColdRateCoefficients}
	\end{figure}

	The computed rate coefficients for an arbitrary value of $M=4$ demonstrates the important role that rotation plays in decreasing charge exchange and increasing proton-impact ionization (\cref{fig:ColdRateCoefficients}).
	
	To give context to the collision rates with supersonic rotation, consider that prior rotating mirror experiments (like MCX \citep{Ellis2005}) produced plasmas with temperatures $\sim10^2$~eV, where the energy losses due to charge exchange actually increase. The target operating temperature of CMFX is $\sim10^3$~eV, where the charge exchange rate is roughly equal for both the rotating and non-rotating plasmas. Lastly, for reactor-scale rotating mirrors $\sim10^4$~eV, charge exchange losses are predicted to decrease in a rotating plasma by several orders of magnitude as evidenced in \cref{fig:ColdRateCoefficients}. Moreover, proton impact ionization increases by several orders of magnitude for all temperatures.
	
	We pessimistically assume that when a charge exchange occurs, it produces a hot neutral at the rotational energy of the plasma that immediately exits the plasma. The mean free path of a hot neutral, $\lambda_{n^*}$, is given by
	\begin{equation}
		\label{eq:mfp}
		\lambda_{n^*} = \frac{v_{n^*}}{n_e \langle \sigma v \rangle_{n^*}} \approx \frac{|\bm{u}|}{n_e \langle \sigma v \rangle_{n^*}} = \frac{M \sqrt{\frac{T_e}{m_i}}}{n_e \langle \sigma v \rangle_{n^*}},
	\end{equation}
	where $v_{n^*}$ is the velocity of a hot neutral, which we assume comes from the kinetic energy of the ions (usually much greater than the thermal energy for rapidly rotating plasmas). And for CMFX- and reactor-relevant plasmas $\lambda_{n^*} \gg a$, so the prompt loss assumption is typically valid. This model does also assume that the neutrals are supplied by an ambient gas puff source, whereas a method like neutral beam injection may be able to decrease charge exchange losses.
	
	In fact, charge exchange can be the dominant mechanism for heat and momentum loss, especially at lower temperatures (see \cref{sec:Power Losses}). An increase in Mach number drastically decreases the charge exchange loss rate at a given temperature, so faster rotation is paramount to decreasing power draw.
	
	\subsection{\label{sec:Radiative Losses}Radiative Losses}
	The \mctrans{} model includes continuum radiation from Bremsstrahlung and cyclotron emission. Bremsstrahlung was modeled with the ``lumped impurity'' assumption and is a function of $Z_{\mathrm{eff}}$, the effective charge of the plasma. However, we assume that the impurity ions do not dilute the main ion species, so quasineutrality is enforced by only considering the main ions and electrons.
	
	Additionally, we assume that the plasma is in coronal equilibrium, meaning that the deexcitation frequency is much larger than the excitation frequency. Therefore, we only consider species in the ground state.
	
	We have assumed that line radiation is negligible, which is true for a sufficiently hot and pure plasma. Future iterations of \mctrans{} may include this effect.
	
	\subsubsection{Bremsstrahlung and Cyclotron Radiation}
	Taking the formula for Bremsstrahlung radiation from the NRL formulary \citep[(62) in][]{nrlformulary} and writing it in convenient units we have
	\begin{equation}
		\dot{Q}_{\mathrm{Brem}} = 5.34 \times 10^3\, Z_{\mathrm{eff}} \left(\frac{n_e}{10^{20} \mathrm{m}^{-3}}\right)^2 \left(\frac{T_e}{1 \mathrm{keV}}\right)^{1/2}\, \mathrm{W}/\mathrm{m}^3,
	\end{equation}
	where we have used the definition
	\begin{equation}
		Z_{\mathrm{eff}} = \frac{1}{n_e} \sum_{s} Z_s^2 n_s
	\end{equation}
	of the effective charge state of the plasma (summation taken over all positively charged species). Similarly, the cyclotron radiation by an electron in vacuum (from the formulary \citep{nrlformulary}) is (in SI units),
	\begin{equation}
		\dot{Q}_{\mathrm{cyc,vac}} = \num{6.21e-17} \ B^2 n_e T_e \ \mathrm{W}/\mathrm{m}^3.
	\end{equation}
	
	The transparency of the plasma to cyclotron emission is calculated from the formulae in \citet{Tamor1983}. We have determined that all plasmas considered here are opaque to cyclotron emission and thus reabsorb most of the radiation. There is some amount of losses that occur at the surface of the plasma, and these are accounted for by assuming a reflection coefficient from the the vacuum vessel of $R= 95$\% (modern vacuum vessels will typically exceed this reflectivity). The total power loss is $\dot{Q}_{\mathrm{cyc}} = k \dot{Q}_{\mathrm{cyc,vac}}$\footnote{The symbol for the transparency factor in \citep{Tamor1983} has been changed from $\Phi$ to $k$ and the factor $\lambda$ to $l$ to avoid confusion with other variables in this work.}, where
	\begin{align}
		k &= \frac{T_e^{3/2}}{200 l^{1/2}},\\
		l &= \frac{2 a}{l_0 (1 - R)},\\
		l_0 &= \num{1.66e16} \ \frac{B}{n_e} \, \mathrm{m},
	\end{align}
	and all quantities are in SI units.
	
	\subsection{\label{sec:Alpha Particles}Alpha Particles}
	
	Our model for alpha particles is that they are all born with a delta-function distribution at the birth energy of $E_\alpha = 3.52$ MeV:
	\begin{equation}
		\left( \pd{f_\alpha}{t}\right)_{\mathrm{Source}} = \frac{S_\alpha \delta(v-v_*)}{4\pi v_*^2},
	\end{equation}
	where $v_*$ is the birth velocity corresponding to $E_\alpha$ and $S_\alpha$ is the birth rate of alphas per unit time per unit volume, retrieved from the Maxwellian-averaged formula in \citet[][p. 45]{nrlformulary}.
	
	Alphas are born at a high energy and lose energy via friction to the electrons. Eventually, they will reach a critical velocity, $v_c$, where they begin to scatter off ions. \citet{Helander2002} gives an approximation of this critical velocity (or in this case, energy) as
	\begin{equation}
		\frac{m_\alpha v_c^2}{2} \sim 50 T_e.
	\end{equation}
	This critical energy is still significantly greater than the centrifugal potential well \cref{eq:ion_potential}, leading to the relation
	\begin{equation}
		\label{eq:alpha_vs_potential}
		\Xi_\alpha \sim \frac{M^2 T_e}{4} \ll \frac{m_\alpha v_c^2}{2} \sim 50 T_e,
	\end{equation}
	as long as $M^2 / 4 \ll 50$ (which generally is true). So we can assume that the alphas are not centrifugally confined and generally do not deposit energy into the ions. However, energy is lost through drag to alpha-electron collisions. The assumption that all the alpha energy is deposited into the electrons is pessimistic because hotter electrons leads to worse ion confinement to maintain quasineutrality.
	
	As alpha particles are born isotropically, a fraction of them are born directly into the unconfined region of velocity space. We thus model the loss region for energetic alphas as a cone, and only classical mirror confinement applies, thus an alpha is lost if
	\begin{equation}
		\mu_\alpha B_{\mathrm{max}} > E_\alpha,
	\end{equation}
	where $\mu_\alpha$ is the magnetic moment. Integrating this over all velocity space, we see that the fraction of alphas that is lost is \citep{ReviewsofPlasmaPhysics13}
	\begin{equation}
		f_{\mathrm{lost}} = 1 - \sqrt{1 - \frac{1}{R_{\mathrm{mirror}}}}.
	\end{equation}
	This fraction of alpha particles is lost promptly and is taken into account in assessing quasineutrality.
	
	Currently, we do not have any explicit collisional losses of alpha particles. However, under the twin assumptions that of $R_{\mathrm{mirror}} \gg 1$ and that we can treat alphas like primary ions, \citep[see][]{Ryutov1988}, the approximate lifetime of alpha particles in the machine is
	\begin{equation}
		\tau_\alpha \approx 0.4 \tau_{\alpha i} \ln R_{\mathrm{mirror}},
	\end{equation}
	where $\tau_{\alpha i}$ is the alpha-ion collision time, as alpha-electron collisions do not scatter the alpha particles \citep{Helander2002}. Additionally, we do not account for dilution of the primary ion species due to the accumulation of helium ash, nor do we account for loss-cone-driven instabilities such as those discussed in \citet{Hanson1984}.	
	
	\section{\label{sec:Results and Discussion}Results and Discussion}
	\mctrans{} can be run in two steady-state modes: a single point in parameter space and batch mode to perform parameter scans. It also offers time-dependent options, including a capacitor bank discharge and `free-wheeling' mode where a steady-state plasma spins down, i.e. is discharged through a dump resistor. The following sections discuss results from all these modes of operation.
	
	\subsection{\label{sec:Experimental Comparison} Experimental Comparison}
	Previously, there have been three centrifugal mirror experiments, including Ixion \citep{Baker1961, Baker1961_2}, PSP-2 \citep{Volosov2009}, and MCX \citep{Ellis2005}. Results from \mctrans{} are compared to the available data of those experiments (\cref{tab:benchmarking}). The following paragraphs detail the assumptions made to model each experiment in \mctrans{}.
	
	\subsubsection{\label{sec:Prior Experiments}Prior Experiments}
	In Ixion \citep{Baker1961}, pre-ionized deuterium gas was injected into the chamber from the side and a negative bias was applied with short inner electrodes via capacitor bank discharges up to 20 kV (though usually 7.5 kV). When the gas was ionized, an axial column of plasma extended from one electrode to the other, thus creating a ``plasma center electrode.'' The diameter of this plasma column at the midplane was $\sim$1.5 times larger than that of the electrode (6.4 cm), and we assume this is the innermost flux surface. The outer radius and axial extent of the plasma was inferred from voltage profile measurements (9.5 cm and 10 cm, respectively). The field at the midplane was 0.95 T, with a mirror ratio of 2.2. \citet{Baker1961} do note that poor vacuum quality could have meant that a typical discharge had as many impurity ions as deuterium, but without an actual measurement, we've assumed $Z_{\mathrm{eff}} = 3$. By modeling the plasma as a capacitor and discharging it into a known resistance, the characteristic ``spin-down'' time can be related to the total charge stored, and therefore the ion density, which was estimated as \num{3e21} m$^{-3}$. Ixion produced highly collisional plasmas (see \cref{tab:operating_conditions}), and because \mctrans{} assumes low-collisionality, the modeled electron temperature (and thus $M_A$) is significantly lower than the measured value. The momentum confinement time, $\tau_M$ is of similar magnitude.
	
	\citet{Volosov2009} wrote a review paper on the PSP-2 experiment which operated from 1975-1985, mostly detailing the work in \citep{Abdrashitov1991}. The electric field was not generated via a center electrode, but rather a series of matching ring electrodes which were charged up to 500 kV (positive bias), though typical discharges were 360 kV. The inner and outer plasma radii are given as 32 and 51 cm, respectively, and the length was measured by neutral detectors to be 40 cm. Hydrogen was simultaneously pumped through six valves spread azimuthally around the midplane. The typical midplane magnetic field was 0.99 T with a mirror ratio of 2.4. Usual discharge densities were on the order of \num{3e17} m$^{-3}$ and \num{1e18} m$^{-3}$ for ionized hydrogen and neutrals, respectively. The relative abundance of impurities is also given and $Z_{\mathrm{eff}}$ was calculated to be $\sim$2.4. The energy confinement time was not calculated, so the ion confinement time ($\tau_D = 100$ \si{\micro\second} in that paper) was used instead. Electron temperatures were indirectly measured based off assumptions relating the ambipolar potential and $T_e$, giving mean electron energies (non-Maxwellian) of 0.1-1 keV. \citep{Abdrashitov1991} reports the mean ion energy was up to 20 keV in the rotating frame; however this was not a direct measurement of ion temperature, rather a measurement of fast neutrals from charge exchange. Ion drift velocities were found to be \num{2e6} m/s. The primary difference between the experimental results and \mctrans{} is in the confinement time. This difference can be primarily attributed to the large neutral inventory (roughly an order-of-magnitude larger than the charged species density). By balancing loss rates and ionization, the neutral source in \mctrans{} is just large enough to keep a constant electron density such that, typically, $n_n \ll n_e$. However, if in reality $n_n \gg n_e$, the ion confinement time will be much smaller due to charge exchange losses.
	
	MCX operated at the University of Maryland until 2012 \citep{Ellis2012}, and \citet{Teodorescu2008,Teodorescu2010} provided an overview of typical experiments. Hydrogen gas was pumped into the chamber until a base pre-fill pressure of $\sim$5 mTorr was achieved. Voltage to a central electrode was applied through a capacitor bank that was typically charged up to -10 kV. The magnetic field was such that $R_{\mathrm{mirror}} = 7.3$ and $B_{\mathrm{min}} = 0.23$ T. The inner conductor and chamber walls limit the inner and outer flux surfaces, at radii of 6 and 26 cm, respectively, with a plasma length of 1.3 m. We again assume $Z_{\mathrm{eff}}=3$. Interferometric methods measured densities in the range of \num{5e20} m$^{-3}$, and thermal electron Bernstein emission provided peak electron temperatures of 100 eV \citep{Reid2014}. \citet{Teodorescu2008} briefly mentions that confinement times of 100 \si{\micro\second} were used for other calculations, as well. The performance of \mctrans{} shows good correspondence to MCX, despite the high collisionality (\cref{tab:benchmarking}).
	
	Many of the experiments described above used a negative (rather than positive) bias to control plasma-surface interactions at the electrodes and aid in breakdown. This effect is only important when the plasma does not yet have a sufficient electron density and conductivity to completely screen imposed parallel electric fields. Once the plasma is fully-ionized and weakly-collisional, the parallel electric field is completely determined by the ambipolar potential from \cref{ambipolar_pot}. At present, \mctrans{} does not model this early phase of the plasma and is thus agnostic to the direction of the electric field.

	\begin{table}
		\begin{center}
			\begin{tblr}{lccccc}
				\textbf{Experiment} & $T_e$ (eV) & $n$ (m$^{-3}$) & $\tau_M$ (\si{\micro\second}) & $M$ & $\nu^*$ \\ \hline
				Ixion (Exp.) & 30 & \num{3e21} & 300 & 1.8 & 815 \\
				Ixion (\mctrans{}) & 3.4 & \num{3e21} & 400 & 0.2 & \num{1.2e3} \\ \hline
				PSP-2 (Exp.) & $\sim$1,000 & \num{3e17} & 100* & 6.2 & \num{1.7e-5} \\
				PSP-2 (\mctrans{}) & 720 & \num{3e17} & \num{9.3e6} & 7.3 & \num{7.5e-5} \\ \hline
				MCX (Exp.) & 100 & \num{5e20} & 100 & 2.6 & 6.3 \\
				MCX (\mctrans{}) & 17 & \num{5e20} & 180 & 6.3 & 112
			\end{tblr}
			\caption{Benchmarking of \mctrans{} against previous experiments. Experimental results for Ixion, PSP-2, and MCX are taken from \citep{Baker1961, Volosov2009, Teodorescu2008}, respectively. Some results did not report momentum confinement time, so *ion confinement time is reported instead. Collisionality is also calculated, and we find that both Ixion and MCX were highly collisional.}
			\label{tab:benchmarking}
		\end{center}
	\end{table}
	
	\subsubsection{\label{sec:CMFX Comparison}CMFX Comparison}
	At the time of writing, CMFX has demonstrated long-lived plasmas with discharges up to 40 kV \citep{Schwartz2023}. Unfortunately, detailed validation between the current model and the experiment is not yet possible because many crucial experimental measurements are still under way -- crucially, $T_i$, $n_e$, and $L$. However, these discharges have provided some results for initial comparison. For this specific section, we will refer to the parameters as ``CMFX (Expt.)'' to avoid confusion with the eventual goals of the experiment, which are referred to as simply ``CMFX'' in the rest of the paper.
	
	The following discussion compares \mctrans{} to preliminary results from CMFX. Capacitor discharges of nominally 25 kV resulted in steady-state deuterium plasmas at $8.1 \pm 0.1$ \si{\kilo \volt}. The voltage and current traces provide estimates of some global variables like resistance, stored energy, capacitance, and momentum confinement time. \cref{tab:cmfx_exp_comp} displays the configuration parameters for \mctrans{} and a comparison to these preliminary experimental results. Due to unavoidable experimental uncertainties and unknowns, we find this result is consistent with our model.
	
	\begin{table}
		\begin{center}
			\begin{tblr}{colspec = {l|l|c|c|},
						 vlines = {2-17}{solid}}
				\cline{2-4}
				& \textbf{Parameter} & \textbf{\mctrans{}} & \textbf{CMFX (Expt.)} \\
				\cline{2-4} \hline
				\SetCell[r=10]{c} Inputs & Fuel & Deuterium & Deuterium \\
				& $B_{\mathrm{max}}$ (\si{\tesla}) & 3.0 & 3.0 \\
				& $B_{\mathrm{min}}$ (\si{\tesla}) & 0.34 & 0.34 \\
				& $n_e$ (\si{\per\meter\cubed}) & \num{2.5e19} & - \\
				& $n_n$ (\si{\per\meter\cubed}) & \num{1e13} & - \\
				& Voltage (\si{\kilo \volt}) & 8.1 & $8.1 \pm 0.1$ \\
				& $2a$ (cm) & 16 & 16 \\
				& $L$ (m) & 0.9 & - \\
				& $Z_{\mathrm{eff}}$ & 3.0 & - \\
				& \texttt{ParallelLossFactor} & 0.1 & - \\
				\hline \hline
				\SetCell[r=6]{c} Outputs & Resistance (\si{\kilo \ohm}) & 8.8 & $9.4 \pm 1.8$ \\
				& Stored Energy (\si{\joule}) & 132 & $133 \pm 5$ \\
				& Capacitance (\si{\micro \farad}) & 3.3 & $4.0 \pm 0.2$ \\
				& Mom. Conf. Time (\si{\milli \second}) & 29 & $38 \pm 7$ \\
				& Strongly-magnetized ($\rho^* = \frac{\rho_i}{a} \ll 1$) & 0.021 & - \\
				& Low-collisionality ($\nu^* = \frac{\nu_{ii} L_\parallel}{\cs} \ll 1$) & 4.6 & - \\
				\hline
			\end{tblr}
			\caption{Nominal configuration parameters and outputs for a CMFX (Expt.) comparison. All the fields with `-' are currently unavailable measurements or not applicable parameters. This comparison to recent experimental results should not be confused with the mention of CMFX in the rest of the paper, which considers the eventual operational goals of the device.}
			\label{tab:cmfx_exp_comp}
		\end{center}
	\end{table}
	
	\mctrans{} provides optional variables to help fit experimental data when values like temperature or density are unavailable. \texttt{ParallelLossFactor} and \texttt{NeutralDensity} ($n_n$) were varied until the results approximately converged on the experimental data. Because \mctrans{} assumes a low-collisionality plasma, it over-predicts loss rates for collisional plasmas; thus, because the predicted value of $\nu^* = 4.6 > 1$, we have set \texttt{ParallelLossFactor} to 0.1, i.e. the parallel loss rates have been artificially modified to 10\% of the value normally predicted by \mctrans{}. Additionally, because gas is only puffed at the beginning of the experiment, not continuously throughout it, we expect the neutral inventory to be quite low. We therefore set $n_n$ to $\sim \num{e-6} n_e$.
	
	\subsection{\label{sec:Scaling}CMFX and Reactor-Scaling}
	To explore the parameter space of interest in CMFX and a reactor, the central field, electron density, and applied voltage were varied. The fixed parameters provided to the input files for \mctrans{} are given in \cref{tab:device_parameters}. The ion density is assumed equal to the electron density to enforce quasineutrality.
	\begin{table}
		\begin{center}
			\SetTblrInner{colsep=3pt}
			\begin{tblr}{lcccccccc}
				\textbf{Device} & Fuel & $B_{max}$ (\si{\tesla}) & $B_{min}$ (\si{\tesla}) & $n_e$ (\si{\per\meter\cubed}) & Voltage (kV) & $2a$ (cm) & $L$ (m) & $Z_{\mathrm{eff}}$ \\ \hline
				\textbf{CMFX} & Hydrogen & 3.0 & 0.34 & \num{1e19} & 100 & 16 & 0.6 & 3.0 \\
				\textbf{Reactor} & DT Fuel & 18.0 & 3.0 & \num{6e19} & 5000 & 25 & 20.0 & 3.0
			\end{tblr}
			\caption{Nominal configuration parameters for CMFX and reactor scenarios considered here.}
			\label{tab:device_parameters}
		\end{center}
	\end{table}
	
	The Alfv\'en speed, $v_A = \infrac{B}{\mu_0 n_i m_i}$, is the speed at which magnetic field perturbations propogate along the axial direction. The best performing plasmas occur when the Alfv\'en Mach number, $M_A \equiv \infrac{u}{v_A}$ approaches unity. However, the results may appear sparse in some areas because configuration parameters with either $M_A > 1.25$ or $\rho^* = \infrac{\rho_i}{a} > 0.1$ were not plotted. As the Alfv\'en Mach number passes unity, the plasma is squeezed in the z-direction into a thin disk by the large centrifugal forces. Eventually, at $M_A \approx 1.25$, the magnetic field no longer has an equilibrium shape as the diamagnetic current and its associated field overcome the vacuum magnetic field, creating a magnetic null and causing the field lines close upon themselves \citep{Abel2022}.
	
	Surpassing the limit $\rho^* \approx 0.1$ brings into question the strongly-magnetized assumption that underlies \cref{parttrans,heattrans,momtrans} and the ordering $\Phi \gg \pot$ \cref{eq:potential_comparison}. However, as opposed to the $M_A$ limit, which results in a plasma with no equilibrium, the $\rho^*$ limit is merely a constraint of the model. Future work should consider the regime where extended gyroradii exist -- in particular the orbits of fusion-produced alpha particles.
	
	Some lower temperature results are limited by the condition that $\nu^* \leq 0.1$. Like the limit for $\rho^*$ (and unlike the $M_A$ limit), this is not a physical limit, but merely a limitation of the model. The assumption that $T_s$ is constant on a flux surface is violated when the plasma becomes collisional, which may affect the results from \cref{sec:Parallel Transport}. Future work should consider plasmas in the collisional regime due to their importance during the start-up phase (Step (iv) in \cref{sec:TimeDependenceTheory}).
	
	The results for the given CMFX and reactor configurations in \cref{tab:device_results} provide some outputs of \mctrans{} for parameters of interest.
	
	\begin{table}
		\begin{center}
			\begin{tabular}{rlcc}
				& \textbf{Parameter} & CMFX & Reactor \\
				\cmidrule{2-4}
				\ldelim\{{4}{*}[Plasma] & $n_e = n_i$ & \num{1e19} \si{\per \meter \cubed} & \num{6e19} \si{\per \meter \cubed} \\
				& $n_n$ & \num{3.9e13} \si{\per \meter \cubed} & \num{2.1e12} \si{\per \meter \cubed} \\
				& $T_i$ & 2.0 keV & 45 keV \\
				& $T_e$ & 1.3 keV & 56 keV \\
				\ldelim\{{8}{*}[System-level] & $P_\mathrm{in}$ & 21 kW & 20 MW \\
				& $P_\mathrm{thermal}$ & - & 170 MW \\
				& $n_i T_i \tau_E$ & \num{2.1e18} \si{\kilo \electronvolt \second \per \meter \cubed} & \num{5.9e21} \si{\kilo \electronvolt \second \per \meter \cubed} \\
				& Neutron rate & - & \num{2.4e19} \si{\per \second} \\
				& Kinetic Energy & 2.2 kJ & 290 MJ \\
				& Thermal Energy & 610 J & 74 MJ \\
				& Resistance & 480 \si{\kilo \ohm} & 1.3 \si{\mega \ohm} \\
				& Capacitance & 440 \si{\pico \farad} & 23 \si{\micro \farad} \\
				\ldelim\{{3}{*}[Momentum loss] & $P_{\parallel i}$ & 510 W & 8.6 MW \\
				& $P_\mathrm{CX}$ & 15 kW & 2.1 MW \\
				\ldelim\{{9}{*}[Heat loss/gain] & $P_{\mathrm{visc}}$ & 5.6 kW & 9.0 MW \\
				& $\dot{Q}_{\mathrm{Brem}}$ & 14 W & 2.2 MW \\
				& $\dot{Q}_{\mathrm{cyc}}$ & 660 \textmu W & 2.6 MW \\
				& $\dot{Q}_{\perp i}$ & 2.2 kW & 2.3 MW \\
				& $\dot{Q}_{\parallel i}$ & 1.1 kW & 7.5 MW \\
				& $\dot{Q}_{\parallel e}$ & 1.4 kW & 19 MW \\
				& $\dot{Q}_\mathrm{CX}$ & 810 W & 81 kW \\
				& $\dot{Q}_\alpha$ & - & 25 MW \\
				\ldelim\{{6}{*}[Confinement time] & $\tau_E$ & 110 ms & 2.2 s \\
				& $\tau_{\parallel}$ & 250 ms & 2.8 s \\
				& $\tau_\perp$ & 160 ms & 13 s \\
				& $\tau_\mathrm{CX}$ & 300 ms & 270 s \\
				& $\tau_\mathrm{ion}$ & 220 ms & 11 s \\
				& $\tau_\alpha$ & - & 5.7 s \\
				\ldelim\{{7}{*}[Dimensionless] & $M$ & 5.3 & 4.5 \\
				& $M_A$ & 0.78 & 1.25 \\
				& $\beta$ & 0.11 & 0.27 \\
				& $\nu^*$ & \num{1.5e-4} & \num{6.8e-5} \\
				& $\rho^*$ & 0.17	 & 0.091 \\
				& $\Omega_i \tau_i$ & \num{2.2e5} & \num{1.8e7} \\
				& $Q_\mathrm{sci}$ & - & 7.0
			\end{tabular}
			\caption{Results predicted by \mctrans{} for CMFX and reactor configurations. See \cref{tab:device_parameters} for the device parameters. Variables denoted by $P$ indicate angular momentum losses, and those by $\dot{Q}$ heat losses. As described in \cref{sec:Power Losses}, viscous heating mediates between angular momentum and heat, in that it is a loss for the former and a gain for the latter -- hence the crossover of two curly braces. Some values are not reported for CMFX because they are only relevant for devices with D-D or D-T fuel. Additionally, some parameters are listed in earlier tables, and are listed again for convenience.}
			\label{tab:device_results}
		\end{center}
	\end{table}
	
	\subsubsection{\label{sec:Performance Parameters}Performance Parameters}
\begin{figure}
		\centering
		\begin{subfigure}{0.48\textwidth}
			\centering
			\captionsetup{justification=centering,font=small}
			\includegraphics[width=\textwidth]{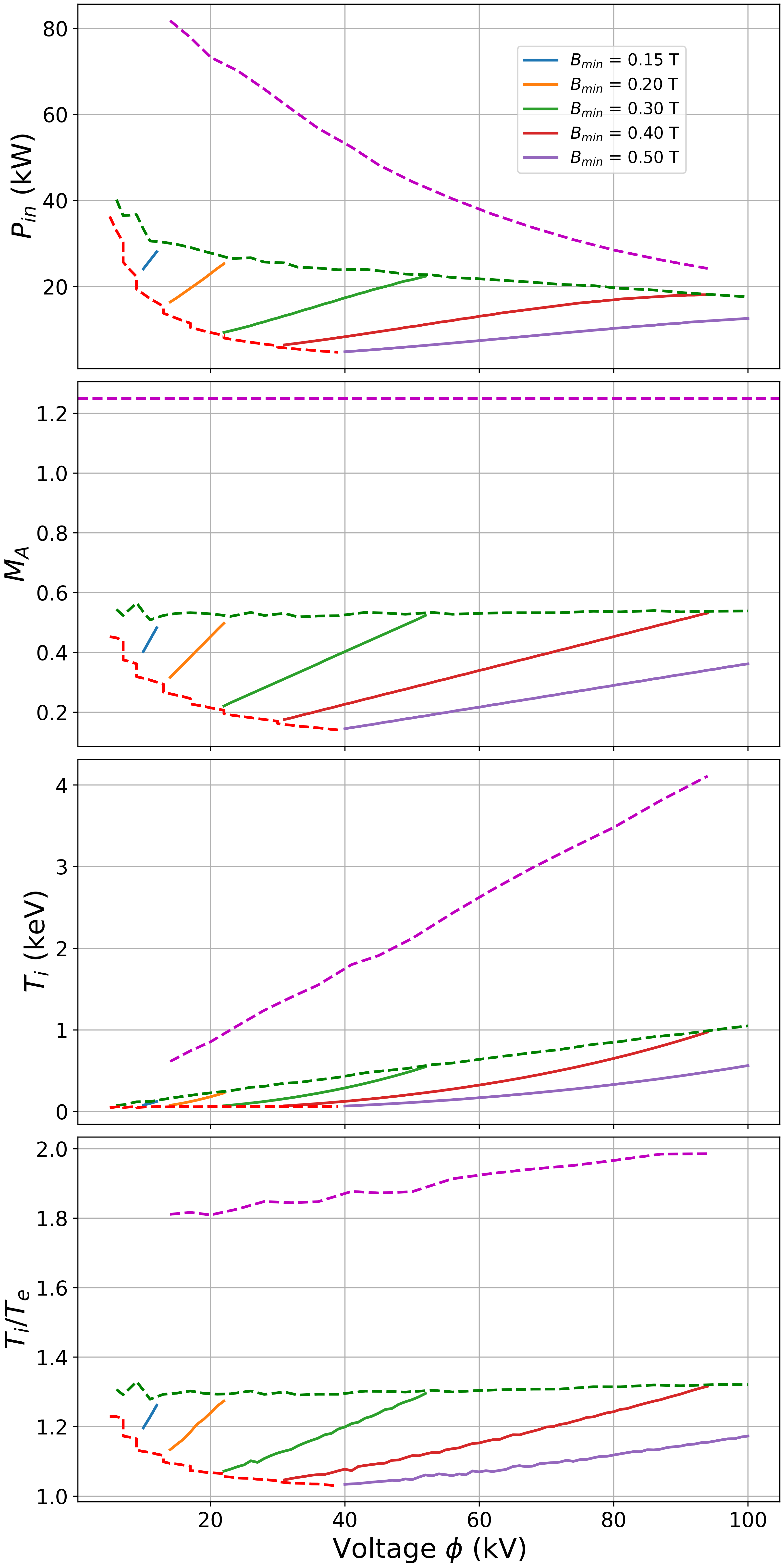}
			\caption{}
			\label{fig:results_CMFX_centralField}
		\end{subfigure}
		\begin{subfigure}{0.48\textwidth}
			\centering
			\captionsetup{justification=centering,font=small}
			\includegraphics[width=\textwidth]{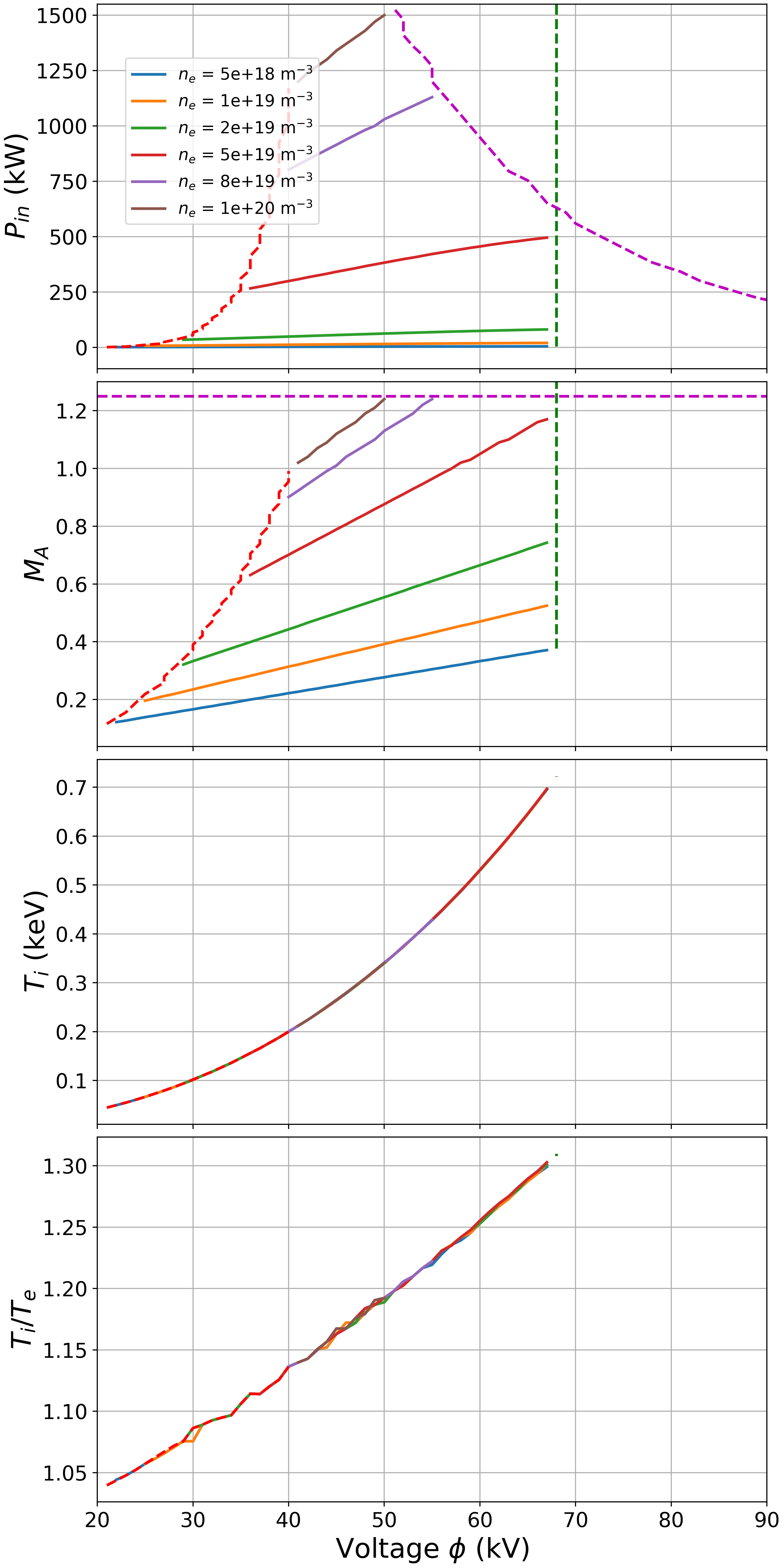}
			\caption{}
			\label{fig:results_CMFX_electronDensity}
		\end{subfigure}
		\caption{Performance of a CMFX-like device for 3 T throat field with (a) a range of central fields ($B_\mathrm{min}$) and fixed electron density ($n_e = \num{1e19}$ m$^{-3}$) and (b) a range of electron densities ($n_e$) and fixed central field ($B_{\mathrm{min}} = 0.3$ T). Results were cutoff above the values of $M_A > 1.25$ and $\rho^* > 0.1$, indicated by the magenta and green dashed lines, respectively.}
		\label{fig:results_CMFX}
	\end{figure}

\begin{figure}
		\centering
		\begin{subfigure}{0.48\textwidth}
			\centering
			\captionsetup{justification=centering,font=small}
			\includegraphics[width=\textwidth]{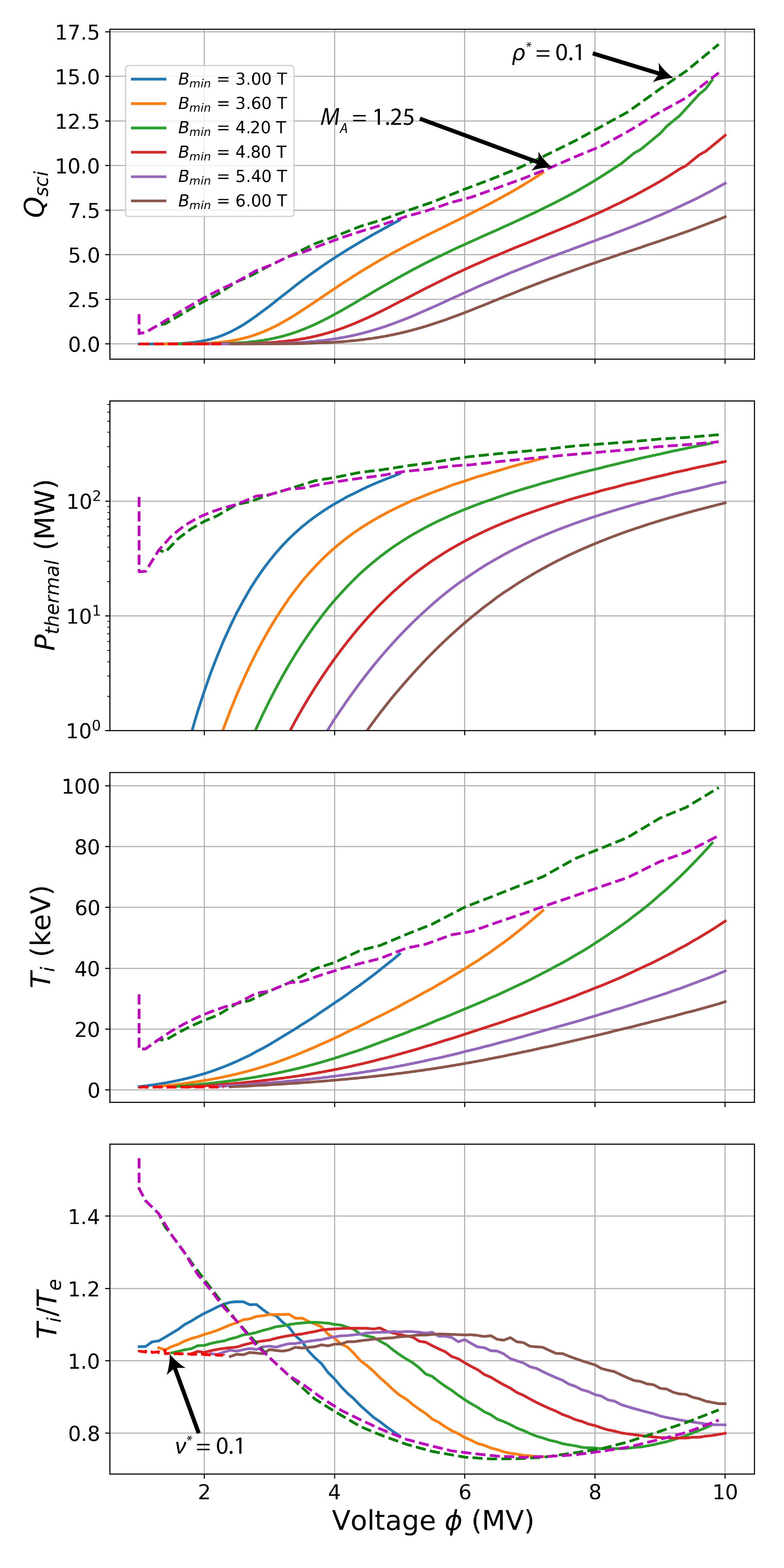}
			\caption{}
			\label{fig:results_reactor_centralField}
		\end{subfigure}
		\begin{subfigure}{0.48\textwidth}
			\centering
			\captionsetup{justification=centering,font=small}
			\includegraphics[width=\textwidth]{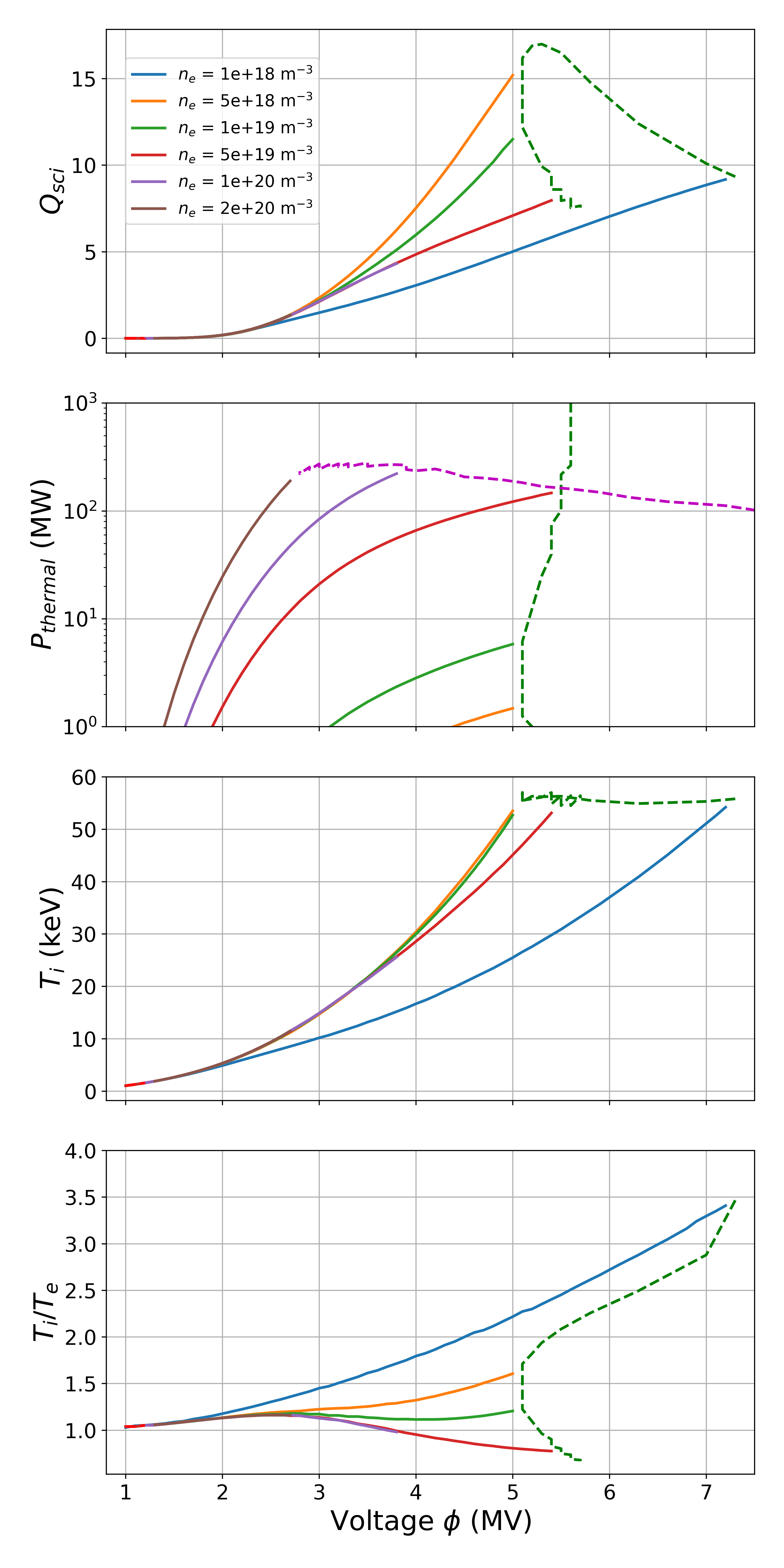}
			\caption{}
			\label{fig:results_reactor_electronDensity}
		\end{subfigure}
		\caption{Performance of a reactor-like device for 18 T throat field with (a) a range of central fields ($B_\mathrm{min}$) and fixed electron density ($n_e = \num{9e19}$ m$^{-3}$) and (b) a range of electron densities ($n_e$) and fixed central field ($B_{\mathrm{min}} = 5$ T). Results were cutoff above the values of $M_A > 1.25$ and $\rho^* > 0.1$, indicated by the magenta and green dashed lines, respectively and where appropriate.}
		\label{fig:results_reactor}
	\end{figure}

	Some parameters of interest were recorded while central field strength, electron density, and voltage were varied. Results for CMFX are in \cref{fig:results_CMFX} and those for a reactor in \cref{fig:results_reactor}. The results were only considered valid if $M_A \leq 1.25$, $\rho^* \leq 0.1$, and $\nu^* \leq 0.1$ as described in the prior section. Dashed lines indicating these limits are in each figure where appropriate.
	
	$P_{\mathrm{in}}$ comes directly from the power supply and is used to drive the rotation. \cref{sec:Power Losses} provides a more in depth explanation of power losses, but generally, because rotational velocity is inversely proportional to $B_{\mathrm{min}}$ ($u \propto E / B$), the power dissipated through viscous torque decreases with larger values of $B_{\mathrm{min}}$. We see that trend reflected in the results for CMFX in \cref{fig:results_CMFX}. There is a trade off in higher rotational velocity, however, between power lost to viscous torque and power saved by enhanced parallel confinement \cref{eq:ParticleLossRate}. For this regime, we see that higher rotational speeds lead to increased power draw.
	
	For the same reason, we see higher ion temperatures for lower values of $B_{\mathrm{min}}$ (or higher values of $\Phi$) because the viscous torque transfers more kinetic (and thus internal) energy to the plasma. We also find that supersonically rotating mirrors organically produce a hot-ion mode. The heating mechanism is viscous shear, which primarily heats the ions because they carry nearly all the rotational momentum in comparison to electrons. Direct heating of the ions is in fact one of the main benefits of supersonic rotation because it negates the need for auxiliary heating systems such as RF or neutral beams.
	
	We also see that CMFX is limited by $\rho^*$, not $M_A$, where the limit scales inversely with $B_{\mathrm{min}}$ and proportionally with $\sqrt{T_i}$ (recall that $\rho^* = \infrac{\sqrt{2 m_i T_i}}{ZeB}$). This limit does not mean that plasmas cannot exist for $\rho^* > 0.1$, but rather the model is not applicable in that regime. We also see that at results are limited at lower temperatures, where $\nu^* > 0.1$ and the plasma is collisional. Lastly, the overlapping lines in \cref{fig:results_CMFX_electronDensity} show that ion and electron temperatures are approximately invariant with density in this temperature regime. 
	
	Similar plots for a reactor scenario (\cref{fig:results_reactor}) demonstrate that the highest performance plasmas are at the limits of $M_A \leq 1.25$ and $\rho^* \leq 0.1$. Moreover, there is competition between high scientific gain, $Q_{\mathrm{sci}}$, and high thermal power output, $P_{\mathrm{thermal}}$. Note this is dissimilar to tokamaks, where high fusion power correlates with high gain \citep{Costley2016}.
	
	Another general observation is that the best performing plasmas operate at ion temperatures that around the peak location of $\sigma_{\mathrm{DT}} \sim 70$ \si{\kilo \electronvolt}. Additionally, we continue to achieve a hot-ion mode at lower voltages, but at higher voltages, a hot-electron mode develops. At these higher voltages, the alpha heating is significantly larger, and we assume that all the alpha energy is deposited into the electrons (see \cref{sec:Alpha Particles}). This is a pessimistic assumption because some amount of alpha energy will be transferred to the ions, with the exact proportion dependent on the precise conditions.
	
	By observing the dashed lines in \cref{fig:results_reactor_centralField}, we see that increasing the central field for a reactor scenario allows for much higher $Q_{\mathrm{sci}}$ (though it does require larger applied voltages). In contrast, the same dashed lines have negative curvature in the graph for $P_{\mathrm{thermal}}$, appearing to reach some asymptotic value. Thus, for a desired $P_{\mathrm{thermal}}$, achieving higher $Q_{\mathrm{sci}}$, requires both larger magnetic field and voltage. The $\nu^*$ limit only affects the lowest temperature discharges in this regime.
	
	The changing density in \cref{fig:results_reactor_electronDensity} is a more complicated story. The performance is primarily limited by $\rho^* \leq 0.1$, except for very high values of density $\gtrsim \num{1e20}$ \si{\per \meter \cubed}. Interestingly, there is a maximum value of $Q_{\mathrm{sci}}$ for $n_e \sim \num{5e18}$ \si{\per \meter \cubed}, but it has quite a low thermal power. And while low densities are not limited by $M_A$ or $\rho^*$, the output power is typically too low because fusion power is proportional to $n_s^2$ \citep{nrlformulary}.
	
	\subsubsection{\label{sec:Power Losses}Power Losses}
	
\begin{figure}
		\centering
		\begin{subfigure}{0.48\textwidth}
			\centering
			\captionsetup{justification=centering,font=small}
			\includegraphics[width=\textwidth]{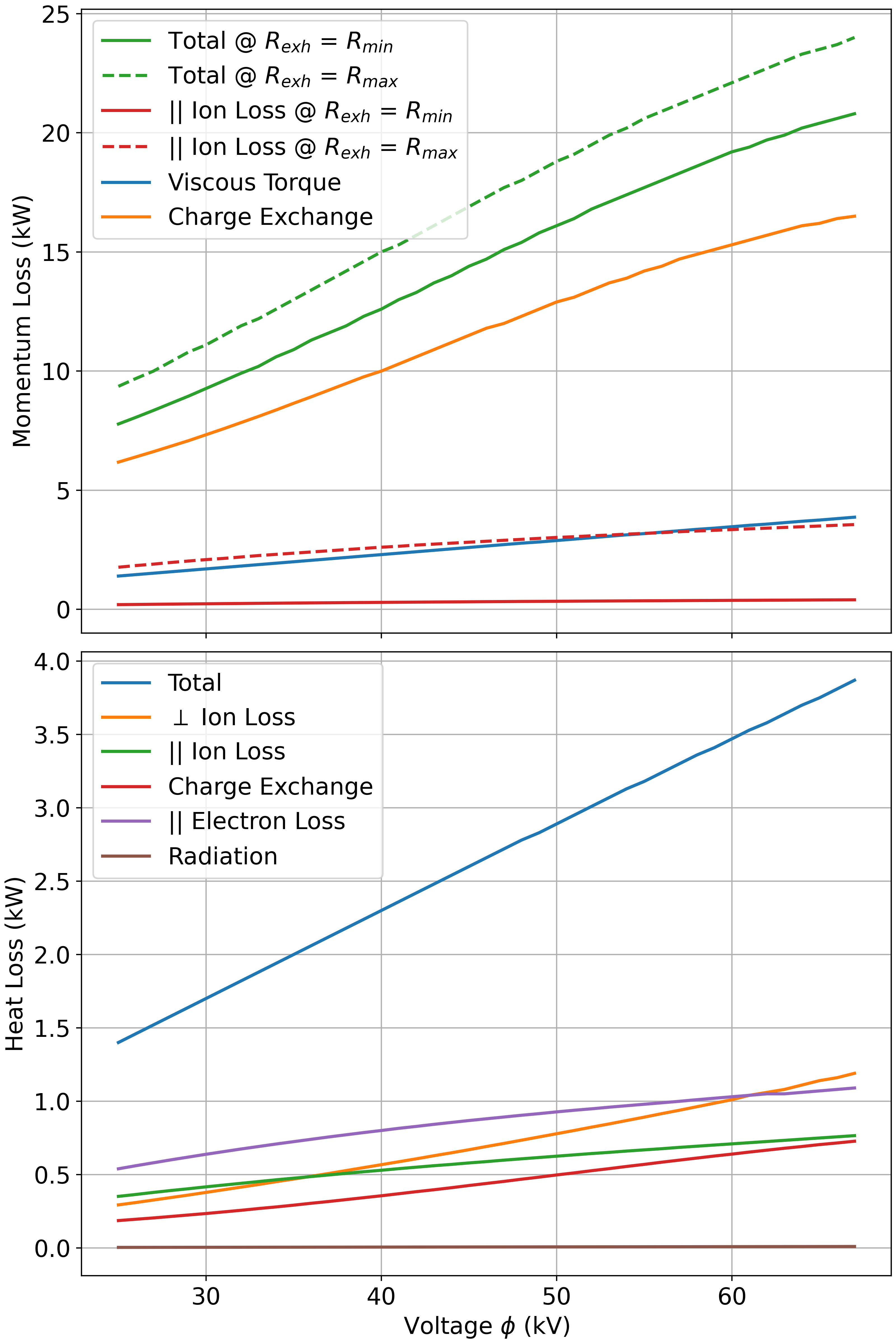}
			\caption{}
			\label{fig:results_losses_CMFX}
		\end{subfigure}
		\begin{subfigure}{0.48\textwidth}
			\centering
			\captionsetup{justification=centering,font=small}
			\includegraphics[width=\textwidth]{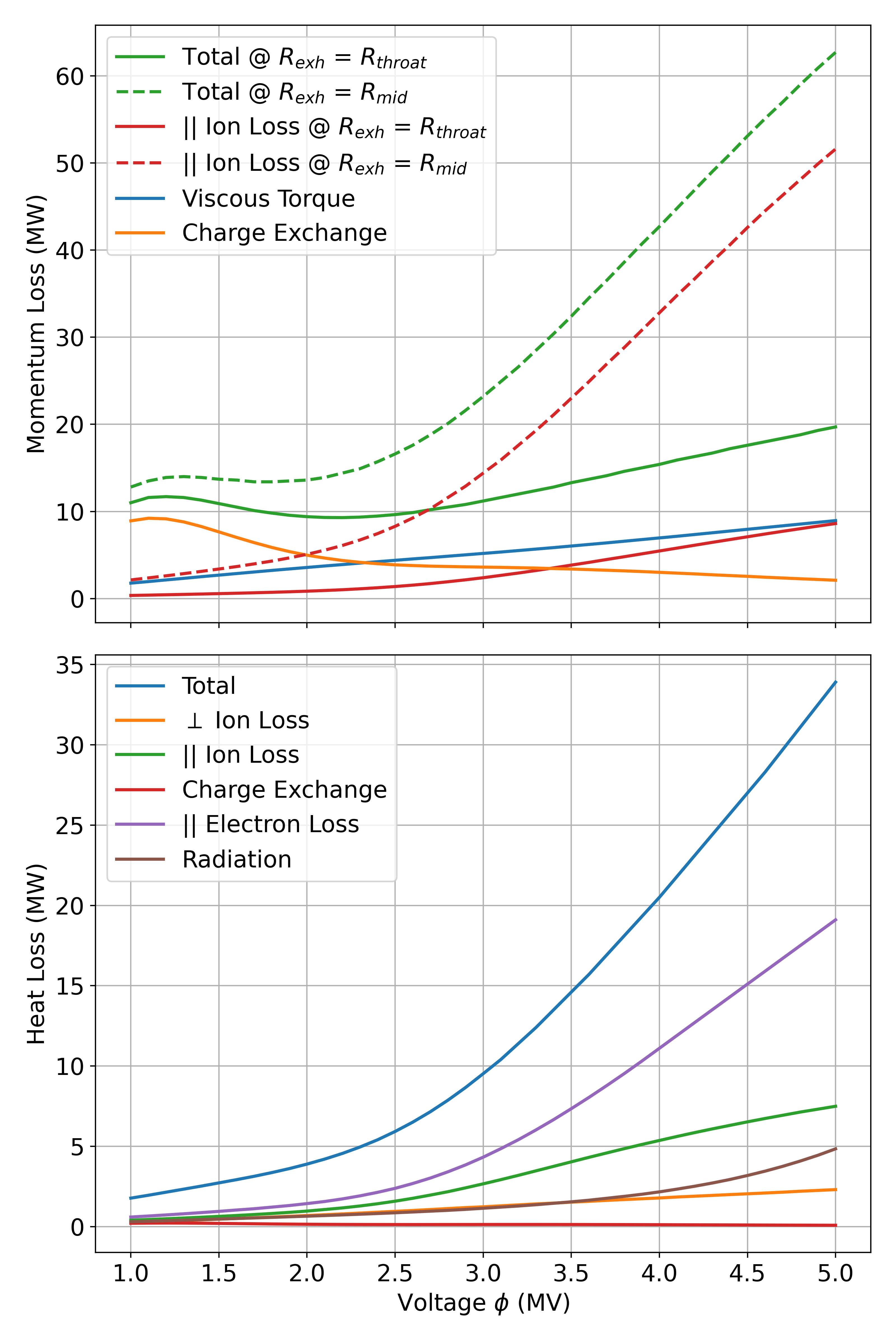}
			\caption{}
			\label{fig:results_losses_reactor}
		\end{subfigure}
		\caption{Power losses for (a) CMFX-like configuration ($n_e = \num{1e19}$ m$^{-3}$ and $B_{\mathrm{min}} = 0.3$ T) and (b) reactor configuration ($n_e = \num{9e19}$ m$^{-3}$ and $B_{\mathrm{min}} = 4$ T). Note the different scales on the horizontal axes for the applied voltage. The momentum lost due to parallel ion losses are a function of $R_\mathrm{exh}$, and the results for $R_\mathrm{exh} = R_\mathrm{min}$ are shown with solid lines, and those for $R_\mathrm{exh} = R_\mathrm{max}$ with dashed lines.}
		\label{fig:results_power}
	\end{figure}

	We assume that a DC power supply directly drives the rotation. The power balance in \cref{momtrans} is used to solve for the radial current density, $j_R$. The current draw from the power supply is balanced by angular momentum losses including particle losses and viscous heating. The viscous heating increases the thermal energy in \cref{heattrans}, where the loss mechanisms are particle losses and radiation. Essentially, viscous heating acts like a mediator between momentum and heat -- a loss term for the prior and a source term for the latter.
	A breakdown of loss mechanisms for both heat and momentum for a CMFX- and reactor scenario shows the importance of each at various applied voltages (\cref{fig:results_power}). As described in \cref{sec:Parallel Losses}, we optimistically choose $R_\mathrm{exh} = R_\mathrm{min}$, which affects the angular momentum lost by a particle in the parallel direction, $m_s \omega R_\mathrm{exh}^2$. For transparency, \cref{fig:results_power} also includes the pessimistic assumption that $R_\mathrm{exh} = R_\mathrm{max}$.
	
	In a CMFX-scenario with lower voltages (corresponding to temperatures $\lesssim$1 keV), charge exchange is the dominant momentum loss mechanism, followed by viscous torque and then parallel ion losses. In the case where $R_\mathrm{exh} = R_\mathrm{max}$, the momentum lost due to ion losses is a factor of $R_\mathrm{mirror}$ larger. The rate coefficient for charge exchange is significantly higher than either electron- or ion-impact ionization (\cref{fig:ColdRateCoefficients}) for a given rotational speed, meaning that the ion loss rate due to charge exchange is large. However, there appears to be a local maximum in momentum losses from charge exchange, indicating that at sufficiently high temperatures, charge exchange ceases to be a limiting factor.
	
	The heat loss mechanisms for CMFX monotonically increase, all being of equal magnitude except for radiation. The heat lost to charge exchange is not the dominant mechanism, as in momentum losses, because the rotational energy of lost ions is much greater than the thermal energy.
	
	For a reactor scenario, viscous torque is the primary momentum loss mechanism, comparable to parallel ion losses at higher voltages. If we take $R_\mathrm{exh} = R_\mathrm{max}$, the momentum loss due to the parallel ion loss rate is dominant. Similarly, as the rotational speed (and therefore temperature) increases, we expect viscous torque to increase and charge exchange to decrease.
	
	Heat loss in a reactor scenario is drastically different from CMFX in that charge exchange and is very small and radiation becomes important. Here, the primary losses are through parallel transport and radiation (roughly equal Bremsstrahlung and cyclotron, see \cref{tab:device_results}), and charge exchange plays almost no role. The heat lost to electrons slightly dominates over parallel ion losses because the electron loss rate is larger and this device operates in a hot-electron mode.
	
\begin{figure}
		\centering
		\begin{subfigure}{0.48\textwidth}
			\centering
			\captionsetup{justification=centering,font=small}
			\includegraphics[width=\textwidth]{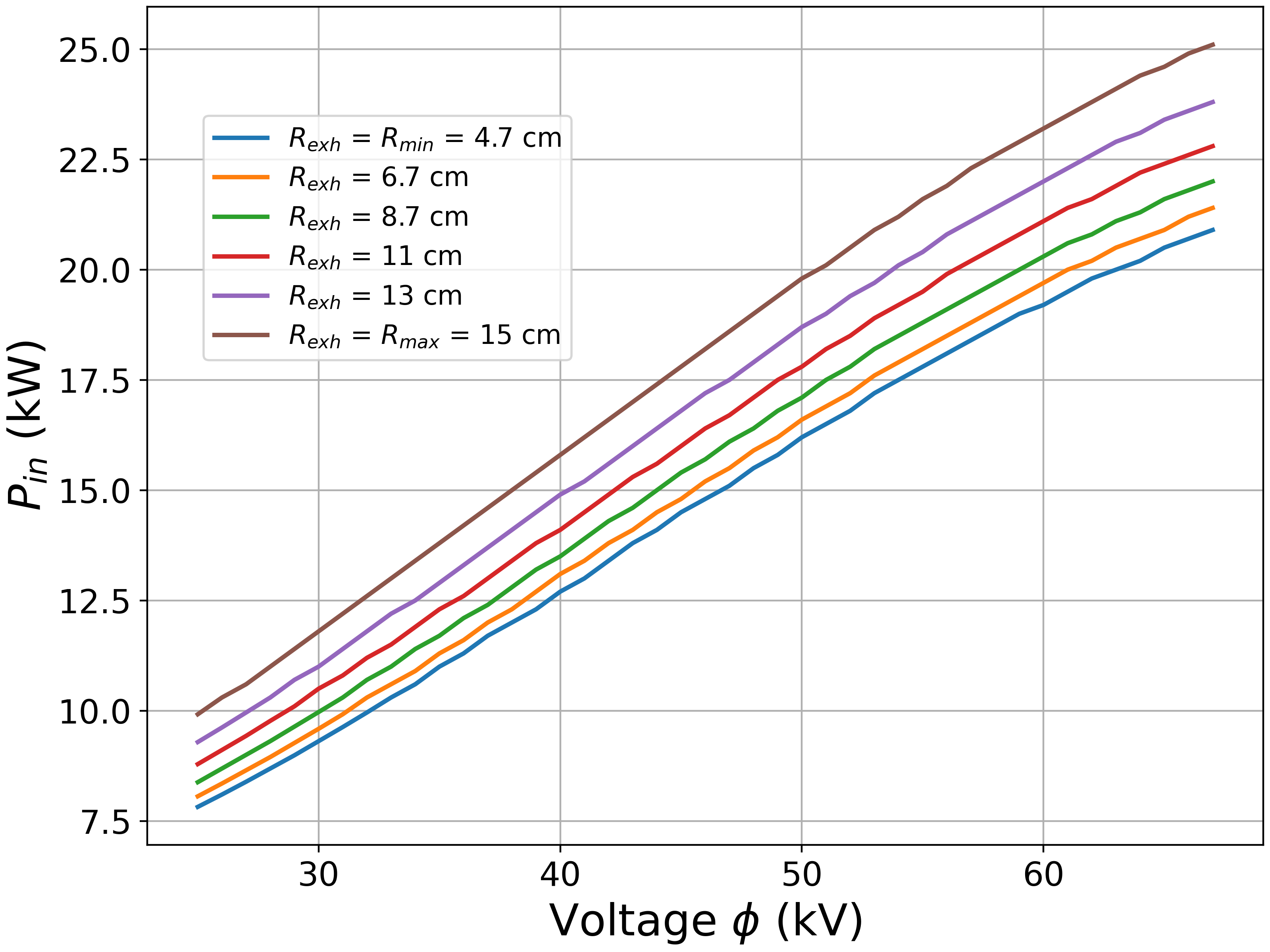}
			\caption{}
			\label{fig:results_CMFX_exhaustRadius}
		\end{subfigure}
		\begin{subfigure}{0.48\textwidth}
			\centering
			\captionsetup{justification=centering,font=small}
			\includegraphics[width=\textwidth]{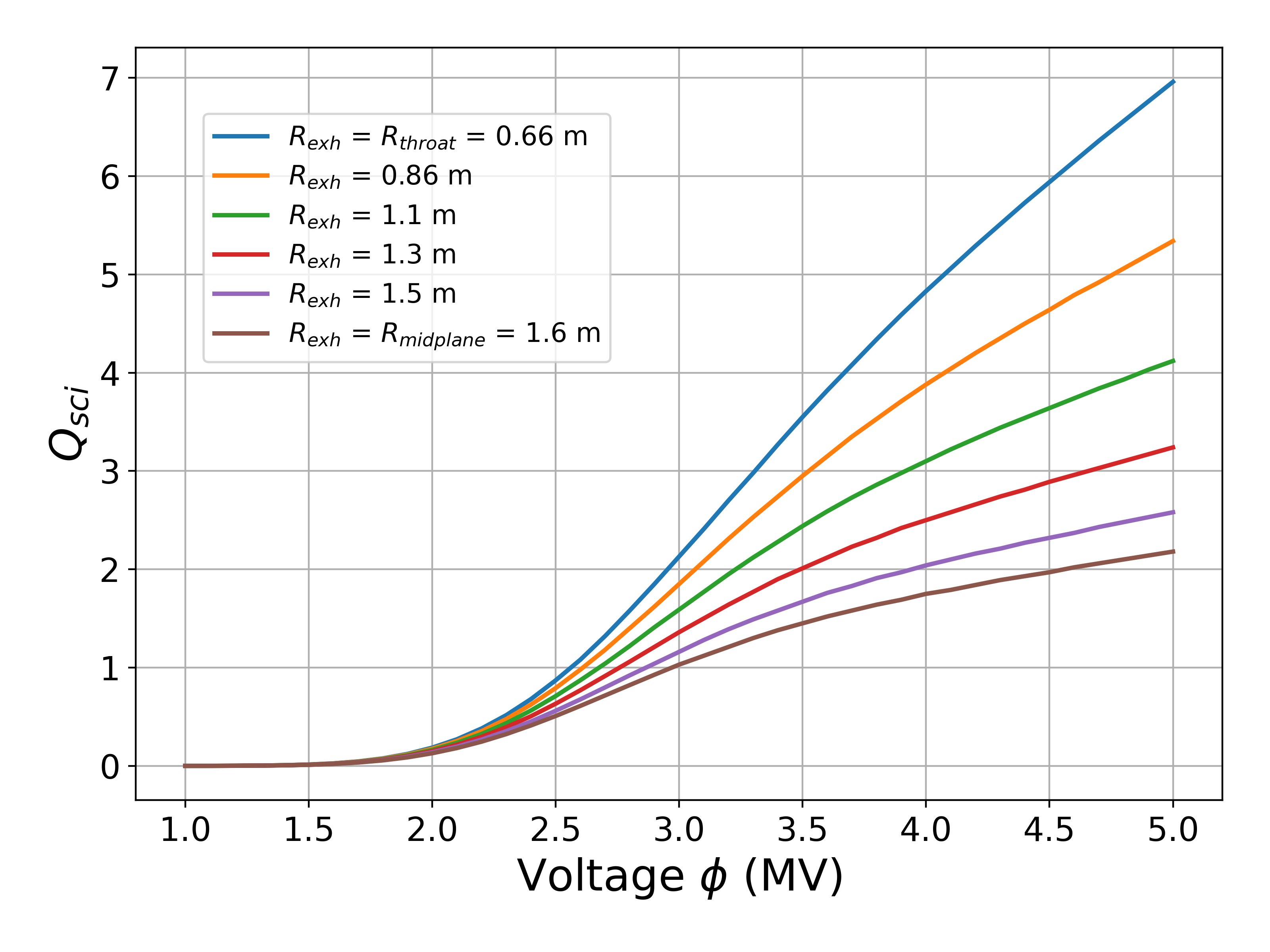}
			\caption{}
			\label{fig:results_reactor_exhaustRadius}
		\end{subfigure}
		\caption{Variation of key parameters for different values of $R_\mathrm{exh}$ for (a) CMFX and (b) reactor scenarios.}
		\label{fig:results_exhaustRadius}
	\end{figure}

	As mentioned, the choice of $R_\mathrm{exh}$ plays a critical role in parallel ion losses for momentum balance. We can see exactly how that choice affects the required $P_\mathrm{in}$ in \cref{fig:results_exhaustRadius}. $P_{\parallel i}$ increases by a factor of $R_\mathrm{mirror}$ from $R_\mathrm{min}$ to $R_\mathrm{max}$. For CMFX, this does not drastically alter the power draw because parallel ion losses are not the dominant loss mechanism of momentum (see \cref{fig:results_power}, charge exchange is dominant). In contrast, $Q_\mathrm{sci}$ in a reactor decreases three-fold because parallel ion losses become dominant with choices of large $R_\mathrm{exh}$.
	
	\subsubsection{\label{sec:Effect of Some Physical Behaviors}Effect of Some Physical Behaviors}
	
\begin{figure}
		\centering
		\begin{subfigure}{0.48\textwidth}
			\centering
			\captionsetup{justification=centering,font=small}
			\includegraphics[width=\textwidth]{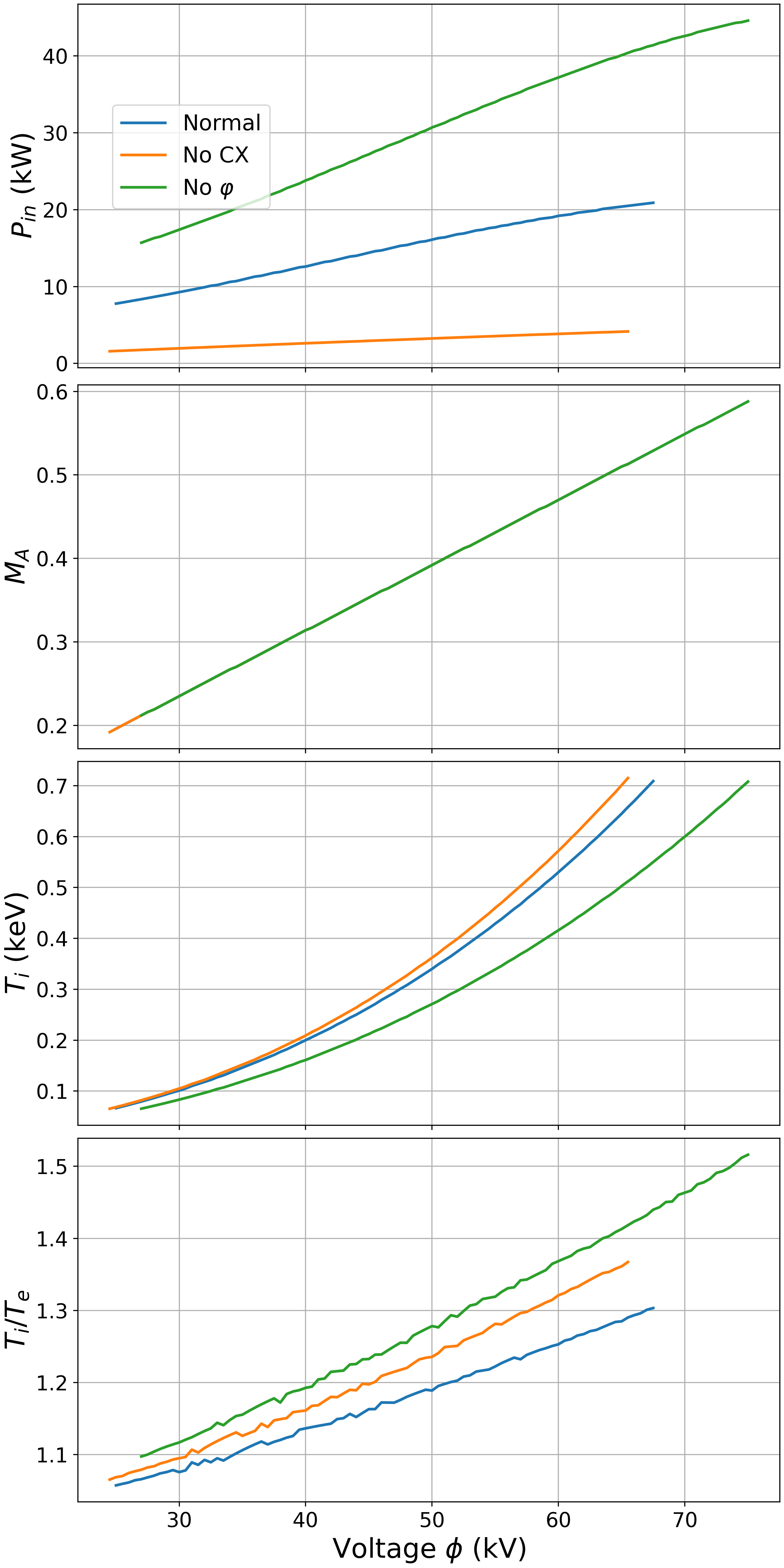}
			\caption{}
			\label{fig:physics_results_CMFX}
		\end{subfigure}
		\begin{subfigure}{0.48\textwidth}
			\centering
			\captionsetup{justification=centering,font=small}
			\includegraphics[width=\textwidth]{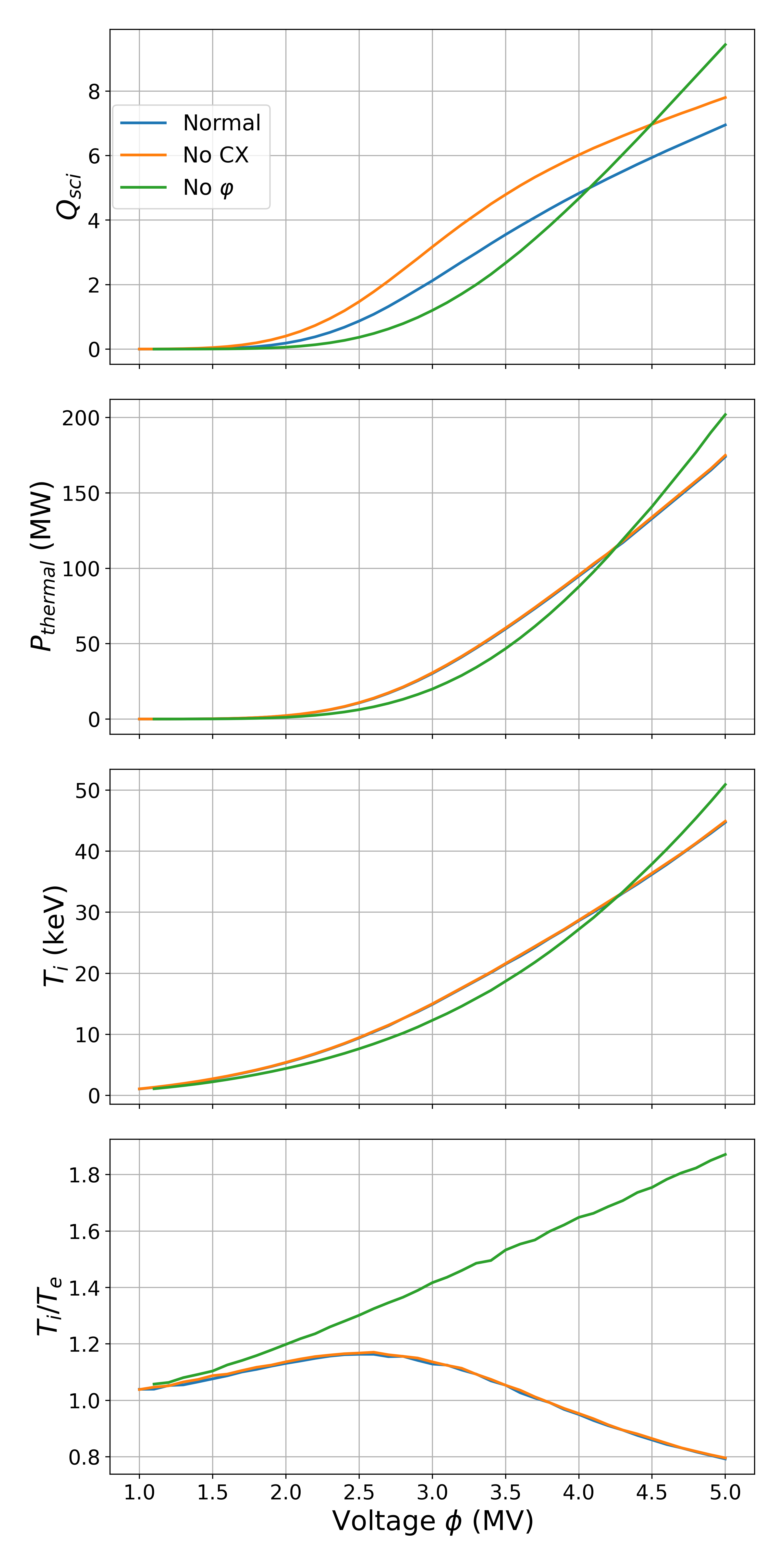}
			\caption{}
			\label{fig:physics_results_reactor}
		\end{subfigure}
		\caption{Effect of turning off charge exchange and ambipolar potential for (a) CMFX-like configuration ($n_e = \num{1e19}$ m$^{-3}$ and $B_{\mathrm{min}} = 0.3$ T) and (b) reactor configuration ($n_e = \num{9e19}$ m$^{-3}$ and $B_{\mathrm{min}} = 4$ T). Note the different scales on the horizontal axes for the applied voltage.}
		\label{fig:physics_results}
	\end{figure}

	\mctrans{} also has the option to turn off some physical behaviors, namely charge exchange and the ambipolar potential. While removing these options is entirely non-physical, the results are useful in demonstrating the importance of charge exchange and the ambipolar potential. The effect of these phenomena can be seen in \cref{fig:physics_results} for both CMFX- and reactor scenarios. In the case where we `turn off' the ambipolar potential, we assume $\pot = \pot_0$, the centrifugal potential. There is then some unbalanced parallel current because we do not consider $\pot_1$, and it is not compensated by a change in radial current. This option is present in the code mainly as a debugging tool. We use it here merely to demonstrate that we do indeed need to consider $\pot_1$ and that its effects are not negligible.
	
	For CMFX, a lack of charge exchange is beneficial because that is the major momentum loss mechanism at these voltages (\cref{fig:results_losses_CMFX}), which in turn increases the ion temperature. Conversely, when we turn off the ambipolar potential, there is a finite axial current (again, non-physical) that leads to higher electron loss rates and lower ion loss rates. Therefore, the viscous torque increases, requiring more power to sustain rotation. Despite the ion temperature decreasing slightly, the electron temperature decreases significantly more, establishing an even more pronounced hot-ion mode.
	
	In contrast, for the reactor scenario, we see that charge exchange plays little role because it is the weakest loss mechanism (\cref{fig:results_losses_reactor}). On the other hand, we see an exaggerated hot-ion mode appear with the absence of $\pot$, leading to a $Q_{\mathrm{sci}}$ value that is quadrupled at some voltages. A lack of $\pot$ allows for a finite parallel current, which produces higher ion confinement. This improved confinement is what sustains a hot-ion mode and therefore high $Q_{\mathrm{sci}}$ values.
	
	\subsection{\label{sec:TimeDependenceResults}Time Dependence}
	
\begin{figure}
		\centering
		\begin{subfigure}{0.48\textwidth}
			\centering
			\captionsetup{justification=centering,font=small}
			\includegraphics[width=\textwidth]{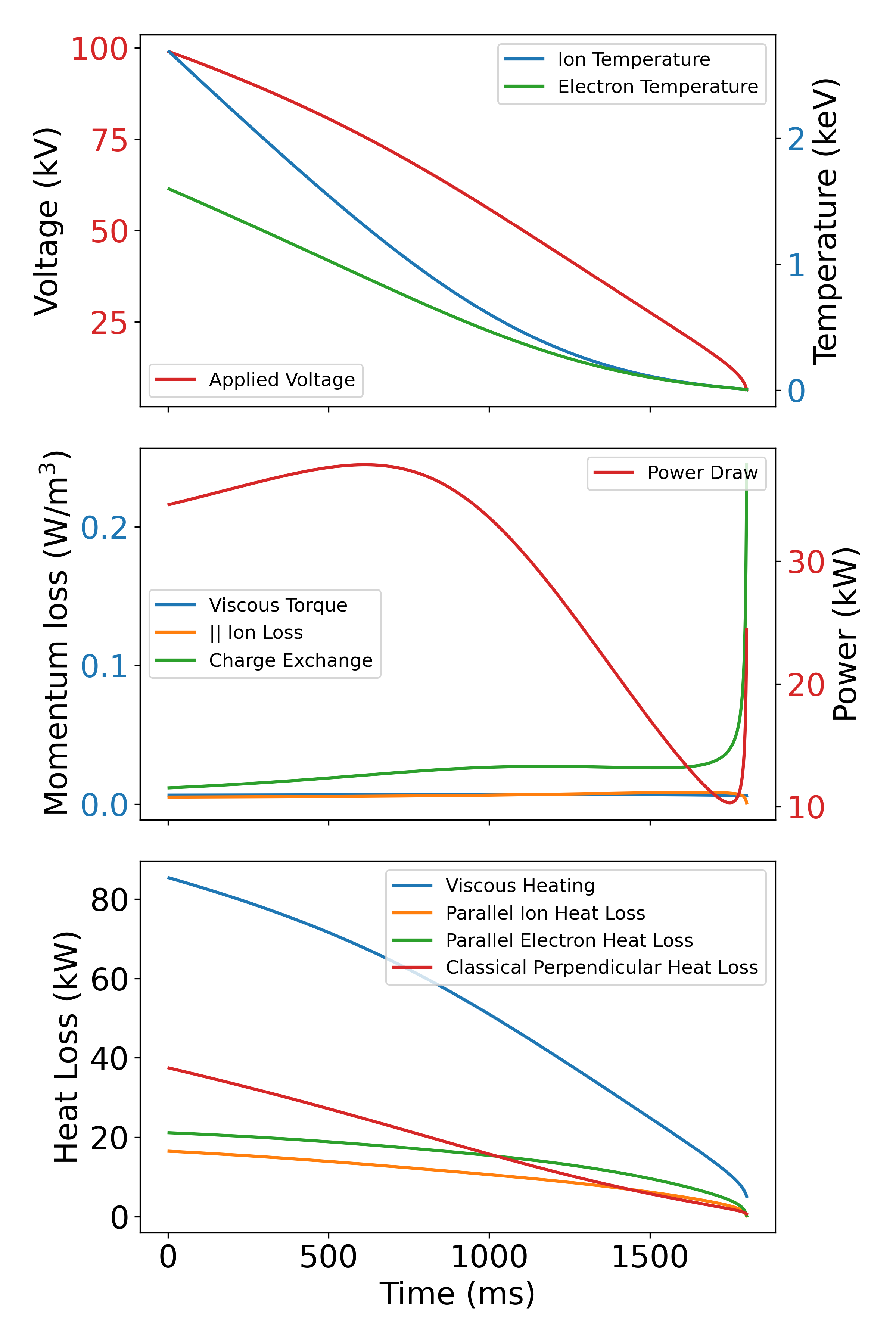}
			\caption{}
			\label{fig:CB}
		\end{subfigure}
		\begin{subfigure}{0.48\textwidth}
			\centering
			\captionsetup{justification=centering,font=small}
			\includegraphics[width=\textwidth]{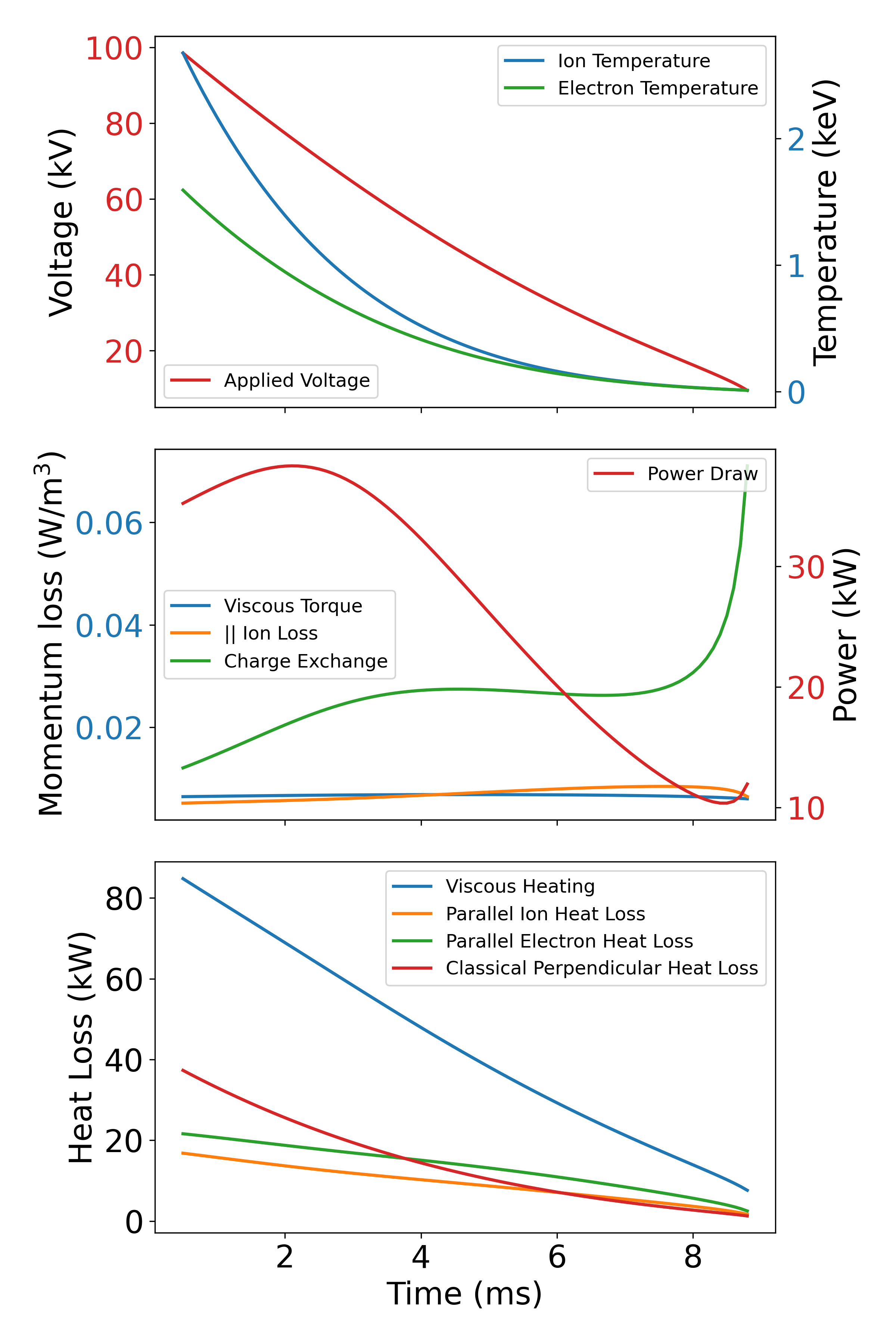}
			\caption{}
			\label{fig:FreeWheelTest}
		\end{subfigure}
		\caption{Time-dependent results for a CMFX configuration. (a) A 72 \si{\micro\farad} capacitor bank, with 1 \si{\giga\ohm} internal resistance, 10 \si{\ohm} line resistance, and 10 \si{\micro\henry} line inductance is discharged into the plasma. (b) At 100 kV, the crowbar is closed and the plasma allowed to discharge into a 10 \si{\kilo\ohm} external resistor. Note the difference in time scales between the two scenarios.}
		\label{fig:time_dependent_results}
	\end{figure}

	As mentioned previously, there are two time-dependent modes of running: (1) capacitor bank discharge and (2) free-wheeling into a load resistor. The 72 \si{\micro\farad} capacitor bank discharge (\cref{fig:CB}) starts with 100 kV across the plasma, and slowly draws current from the capacitors which decrease in voltage. The power draw is extremely non-linear, with a large spike at the end due to an increase in current across the plasma at lower temperatures. The time scale of this behavior is on the order of seconds because the plasma resistance is between \num{10} \si{\kilo\ohm} and \num{100} \si{\kilo\ohm} and therefore the RC time for a capacitor bank of this size is seconds long. 
	
	The free-wheeling mode discharges a CMFX-like plasma at 100 kV into a 10 \si{\kilo\ohm} external dump resistor. The plasma stores energy like a hydromagnetic capacitor, meaning that its behavior will be similar to a capacitor being discharged into the same load. The main difference is that the capacitance of the plasma is variable, and therefore the power draw does not fall exponentially as with a static capacitor. As the voltage, and therefore temperature, drops, charge exchange becomes the dominant loss mechanism. \mctrans{} predicts a plasma capacitance of 440 \si{\pico\farad} for a CMFX-like configuration at 100 kV and a density of \num{1e19} \si{\per\meter\cubed}. With a dump resistance of 10 \si{\kilo\ohm}, the RC time for this system is 4.4 ms. In reality, the voltage decay is slightly faster because the capacitance decreases as the plasma temperature falls.
	
	\section{\label{sec:Conclusions}Conclusions}
	
	\mctrans{} is a 0-D model for centrifugal mirrors. It is based off several assumptions, including a strongly-magnetized and low-collisionality plasma, a large mirror ratio, and supersonic rotation. The confining potential is due to both an ambipolar and centrifugal potential, the latter of which is confining for both ions and electrons. In the low-collisionality limit, perpendicular losses are classical, and continuum radiation losses like Bremsstrahlung and cyclotron emission are included. A neutrals model has been implemented to determine the electron- and ion-impact ionization and charge exchange rates. The retained alpha particles are assumed to deposit all their energy into the electrons, and alphas are otherwise lost through a axisymmetric loss cone.

	Comparison with prior results shows good agreement with MCX. \mctrans{} could not compute a solution for an Ixion-like configuration because the plasma was highly collisional. The comparison with PSP-2 showed differences likely due to the high neutral densities in that device.
	
	Scaling of CMFX demonstrates that it is limited by $M_A$ at higher densities and increased performance as voltage increases. Additionally, supersonic rotation leads to a naturally occurring hot-ion mode, with direct ion heating resulting through momentum transfer to the ions via viscous shear. Scaling of the reactor scenario posits that a balance between low-field and high-density generates high $Q_{\mathrm{sci}}$ and $P_{\mathrm{thermal}}$. A hot-electron mode does appear, but this is due to the alpha particles depositing their energy into the electrons.
	
	We find that at lower ion temperatures, the dominant power loss mechanism is through charge exchange, but at fusion-relevant temperatures, parallel losses become more important. Similarly, we see that radiation heat losses become more important at higher temperatures.
	
	\mctrans{} can simulate time-resolved solutions, including a capacitor discharge and free-wheeling into an external dump resistor. The plasma behaves similarly to a capacitor, storing its energy in rotation. In fact, the time-scales of parameter evolution are very close to the those of an RC circuit. In both cases, as the plasma temperature drops, we see a large spike in power draw due to charge exchange, while all the other loss mechanisms decrease.
	
	Future work will implement line-radiation and collisional transport, as well as model the expander region near the insulators. Work is currently in progress to attain radial profiles and model magnetic equilibrium conditions.
	
	\section*{Funding}
	This work was supported by ARPA-E award no. DE-AR0001270.
	
	\section*{Declaration of Interests}
	The authors report no conflict of interest.
	
	\appendix	
	
	\section{\label{sec:ShearFlowStabilization}Review of Shear Flow Stabilization}
	Shear flow stabilization of turbulence is key to the success of centrifugally confined mirrors. The assumption that perpendicular transport is classical hinges on this behavior. While this phenomenon has been demonstrated for rotating mirrors both theoretically and experimentally, the concept is also pervasive in tokamak literature.
	
	There are two primary instabilities which must be suppressed in order to eliminate turbulent transport. The interchange (also known as ``flute'') instability occurs when the magnetic field curvature and pressure gradients point in the same direction, and is similar to the fluid Rayleigh-Taylor instability, where the effective outward gravity comes from centrifugal rotation \citep{Goldston2020}. The interchange instability occurs on the same scale as the pressure gradient ($\nabla p / p$). This instability was the primary one observed on older mirror machines \citep{Post1987}, which led to the creation of the ``minimum-B'' configuration that included complicated magnetic coil shapes\footnote{Complex coil configurations like the baseball and yin-yang were created to create a magnetic field that had opposite curvature to the pressure gradient. The Tandem Mirror Experiment took this design a step further to utilize the ion-deconfining ambipolar potential and increase $Q_\mathrm{sci}$. While improvement was demonstrated, these devices could not practically be used for power production.} to combat the poor magnetic curvature.
	
	The Kelvin-Helmholtz (K-H) instability can occur when there is shear flow and thus relative motion between fluid surfaces. For a strongly-magnetized plasma, the scale of this instability is much larger than the ion gyroradius \citep{Horton1987}. The concern in a centrifugal mirror is that the shear flow between radial layers might cause K-H vortices to form.
	
	\subsection{\label{sec:ShearFlowTheory}Theoretical}
	\citet{Burrell1997} provides a review of the theoretical underpinnings of shear flow stabilization. A highly idealized approximation demonstrates nonlinear decorrelation of turbulence, whereby radial turbulent flux can be reduced, even if the plasma starts in a highly turbulent configuration. A more complex analysis of linear stabilization is mode-dependent, but the general theme is that unstable modes are coupled to stable modes through velocity shear, which in turn dampen the unstable modes. The key takeaway is that, if the turbulent decorrelation time is long compared to the shearing rate, turbulence is suppressed.
	
	Some work has been specifically applied to stabilization of centrifugally confined plasmas. For example, \citet{Huang2001} used a 3D ideal MHD model to demonstrate that supersonic shear flows suppress growth of the interchange instability. \citet{Huang2004} considered four primary instabilities -- interchange and K-H in the low $\beta$ limit, and magnetorotational and Parker in the high $\beta$ limit. The conclusions for each are as follows: (1) the interchange mode can be suppressed with sufficiently high $M$, the limit of which can be lowered for elongated plasmas, (2) the K-H instability is marginally MHD stable provided there are no inflection points in the density and rotation profile. Particle sources play a key role in attaining stability for these two, and must be considered further. (3) Because $M_A \lesssim 1$, the magnetorotational instability does not occur, and (4) the Parker instability may simply create new equilibria, but further works needs to be done.
	
	Theoretical work on PSP-2 posits that the combination of shear flow and line-tying (achieved by concentric conducting rings to set the electric field profile) are necessary for MHD stability \citep{Volosov2009}. However, that hypothesis is only true for particular radial electric field profiles. The PSP-2 profile is relatively flat (see Fig. 14 in that work), and at any point where the gradient is near zero, there will be no shear and therefore no shear stabilization. In contrast, a singly-peaked profile (as has been achieved with a center conductor \citep{Romero-Talamas2012,Huang2001}) and sufficiently large rotational shear, can stabilize centrifugally-confined plasmas in the absence of line-tying. \citet{Ryutov2011} also states that the centrifugal force alone is sufficient to achieve stability.
	
	Recent theoretical work on shear flow stabilization in tokamaks has demonstrated suppression of all linear turbulence in the condition of zero magnetic shear \citep{Highcock2010}. And large enough shear flow can suppress nonlinear turbulence due to ion temperature gradients (ITG) \citep{Highcock2010,Highcock2011}. For subsonic shear flows, heat is transported neoclassically and momentum by turbulence. However, if the flow is supersonic, the parallel velocity gradient (PVG) instability can cause a transition back to turbulence \citep{Newton2010,Highcock2011}. \citet{Parra2011} concludes that, for a given tokamak configuration, there is an optimal amount of perpendicular momentum injection to suppress turbulence. While tokamak plasmas may become turbulent when shear flow parallel to the magnetic field is too large, the centrifugal mirror does not have this problem because the flow is strictly perpendicular. 
	
	\subsection{\label{sec:ShearFlowExperimental}Experimental}
	\citet{Burrell1997} also includes an extensive review of experimental results that demonstrate shear flow suppression of turbulence. This work only considers tokamaks, but lists experiments like JET and TFTR that demonstrate neoclassical diffusion in the core with shear flow, and like DIII-D and JT-60U that show similar behavior across the entire minor radius. Experiments in JET have demonstrated that higher ITGs can be achieved with larger shear flows, and show good correlation to several gyrokinetic codes \citep{Mantica2009}. More recent work on DIII-D has demonstrated access to a stable `Super-H' mode with large stored energy, primarily due to sheared flow \citep{Knolker2021}.
	
	The review from \citet{Volosov2009} also provides a discussion of experimental results, which show good agreement with the MHD theory underpinning shear flow stabilization in centrifugal mirrors. Results from MCX demonstrated suppression of the interchange instability with rotational speeds an order of magnitude faster than interchange growth times \citep{Ghosh2006}.
	
	\bibliographystyle{jpp}

	\bibliography{references}
	
\end{document}